\documentclass[twocolumn]{aastex631}

\usepackage[utf8]{inputenc}
\usepackage[T1]{fontenc}
\usepackage{amsmath}
\usepackage{graphicx}

\renewcommand{\vec}[1]{\boldsymbol{#1}}

\defcitealias{garavito-camargo_hunting_2019}{GC19}

\received{May 6, 2023}
\revised{June 29, 2023}
\accepted{July 7, 2023}

\shorttitle{The LMC's Dynamical Friction Wake in CDM vs. FDM}
\shortauthors{Foote et al.}
\begin{document}

\title{Structure, Kinematics, and Observability of the Large Magellanic Cloud's Dynamical Friction Wake in Cold vs. Fuzzy Dark Matter}

\correspondingauthor{Hayden R. Foote}
\email{haydenfoote@arizona.edu}

\author[0000-0003-1183-701X]{Hayden R. Foote}
\affil{Steward Observatory, The University of Arizona, 933 North Cherry Avenue, Tucson, AZ 85721, USA.}

\author[0000-0003-0715-2173]{Gurtina Besla}
\affil{Steward Observatory, The University of Arizona, 933 North Cherry Avenue, Tucson, AZ 85721, USA.}

\author[0000-0001-6631-2566]{Philip Mocz}
\affil{Department of Astrophysical Sciences, Princeton University, Princeton, NJ, 08544, USA}
\affil{Lawrence Livermore National Laboratory, 7000 East Ave, Livermore, CA 94550, USA}

\author[0000-0001-7107-1744]{Nicol\'as Garavito-Camargo}
\affil{Center for Computational Astrophysics, Flatiron Institute, Simons Foundation, 162 Fifth Avenue, New York, NY 10010, USA}

\author[0000-0002-0041-4356]{Lachlan Lancaster}
\affil{Department of Astrophysical Sciences, Princeton University, Princeton, NJ, 08544, USA}
\affil{Department of Astronomy, Columbia University, 550 West 120th Street, New York, NY, 10027, USA}
\affil{Center for Computational Astrophysics, Flatiron Institute, Simons Foundation, 162 Fifth Avenue, New York, NY 10010, USA}

\author[0000-0002-9735-3851]{Martin Sparre}
\affil{Institut f\"ur Physik und Astronomie, Universit\"at Potsdam, Karl-Liebknecht-Str 24/25, D-14476 Golm, Germany}
\affil{Leibniz-Institut f\"ur Astrophysik Potsdam (AIP), An der Sternwarte 16, D-14482 Potsdam, Germany}

\author[0000-0002-6993-0826]{Emily C. Cunningham}
\altaffiliation{NASA Hubble Fellow}
\affiliation{Department of Astronomy, Columbia University, 550 West 120th Street, New York, NY, 10027, USA}
\affiliation{Center for Computational Astrophysics, Flatiron Institute, Simons Foundation, 162 Fifth Avenue, New York, NY 10010, USA}

\author[0000-0001-8593-7692]{Mark Vogelsberger}
\affil{Department of Physics, Kavli Institute for Astrophysics and Space Research, Massachusetts Institute of Technology, Cambridge, MA 02139, USA}

\author[0000-0002-1947-333X]{Facundo A. G\'omez}
\affil{Instituto Multidisciplinario de Investigaci\'on y Postgrado,  Universidad de La Serena, La Serena, Chile}
\affil{Departamento de Astronom\'ia, Universidad de La Serena, Av. Juan Cisternas 1200 Norte, La Serena, Chile}

\author[0000-0003-3922-7336]{Chervin F. P. Laporte}
\affil{Departament de F\'isica Qu\`antica i Astrof\'isica (FQA), Universitat de Barcelona (UB), c. Mart\'i i Franqu\`es, 1, 08028 Barcelona, Spain}
\affil{Institut de Ci`encies del Cosmos (ICCUB), Universitat de Barcelona (UB), c. Mart\'i i Franqu\`es, 1, 08028 Barcelona, Spain} 
\affil{Institut d’Estudis Espacials de Catalunya (IEEC), c. Gran Capit\`a, 2-4, 08034 Barcelona, Spain}



\begin{abstract}

The Large Magellanic Cloud (LMC) will induce a dynamical friction (DF) wake on infall to the Milky Way (MW). The MW's stellar halo will respond to the gravity of the LMC and the dark matter (DM) wake, forming a stellar counterpart to the DM wake. This provides a novel opportunity to constrain the properties of the DM particle. We present a suite of high-resolution, windtunnel-style simulations of the LMC’s DF wake that compare the structure, kinematics, and stellar tracer response of the DM wake in cold DM (CDM), with and without self-gravity, vs. fuzzy DM (FDM) with $m_a = 10^{-23}$ eV. We conclude that the self-gravity of the DM wake cannot be ignored. Its inclusion raises the wake's density by $\sim 10\%$, and holds the wake together over larger distances ($\sim$ 50 kpc) than if self-gravity is ignored. The DM wake's mass is comparable to the LMC's infall mass, meaning the DM wake is a significant perturber to the dynamics of MW halo tracers. An FDM wake is more granular in structure and is $\sim 20\%$ dynamically colder than a CDM wake, but with comparable density. The granularity of an FDM wake increases the stars' kinematic response at the percent level compared to CDM, providing a possible avenue of distinguishing a CDM vs. FDM wake. This underscores the need for kinematic measurements of stars in the stellar halo at distances of 70-100 kpc. 

\end{abstract}



\section{Introduction} \label{sec:intro}

The Large Magellanic Cloud (LMC) is the Milky Way's (MW) largest satellite galaxy, possessing an infall mass of $\sim 1$-$2 \times 10^{11}$ M$_\odot$, roughly 10\% that of the MW \citep{besla_simulations_2010, peebles_orbit_2010, penarrubia_timing_2016, patel_orbits_2017, erkal_total_2019, erkal_limit_2020, shipp_measuring_2021, vasiliev_tango_2021, correa_magnus_measuring_2022, koposov_s_2023}. The LMC is currently on its first infall to the MW \citep{besla_are_2007, kallivayalil_third-epoch_2013}, and is inducing significant perturbations in the MW's dark matter (DM) halo, including the collective response, MW reflex motion about the MW/LMC barycenter, and a dynamical friction (DF) wake (\citealt{weinberg_fluctuations_1998, gomez_and_2015, garavito-camargo_hunting_2019}, hereafter \citetalias{garavito-camargo_hunting_2019}; \citealt{garavito-camargo_quantifying_2021, tamfal_revisiting_2021, rozier_constraining_2022}, see also \citealt{vasiliev_effect_2023} for a recent review). 

The perturbations to the MW halo potential induced by the LMC have important, widespread effects on the kinematics of halo tracers, including stellar streams \citep[e.g.][]{vera-ciro_constraints_2013, gomez_and_2015, erkal_modelling_2018, erkal_total_2019, shipp_proper_2019, vasiliev_tango_2021, koposov_s_2023, lilleengen_effect_2023}, globular clusters and satellite galaxies \citep[e.g.][]{erkal_equilibrium_2020, garrow_effects_2020, patel_orbital_2020, correa_magnus_measuring_2022, trelles_concurrent_2022}, and the halo stars in general (e.g. \citetalias{garavito-camargo_hunting_2019}; \citealt{petersen_reflex_2020, cunningham_quantifying_2020, erkal_equilibrium_2020, petersen_detection_2021}). The LMC's infall also affects mass measurements of the MW \citep[e.g.][]{erkal_equilibrium_2020, chamberlain_implications_2023, correa_magnus_measuring_2022} and even the shape and dynamics of the MW's stellar disk \citep[e.g.][]{weinberg_dynamics_1998, laporte_response_2018, laporte_influence_2018}.

If the MW halo's response to the LMC depends on the microphysics of the DM particle, then this scenario presents a unique opportunity to constrain the nature of DM. In particular, the LMC's DF wake offers a promising test-bed, as the strength and density structure of DF wakes depends on the physics of the medium in which they form \citep[e.g.][]{ostriker_dynamical_1999, furlanetto_constraining_2002, lancaster_dynamical_2020, vitsos_dynamical_2023}. However, our limited ability to disentangle the response of halo tracers due to the wake specifically vs. other perturbations induced by the LMC (e.g. the LMC's tidal field and the MW's reflex motion) presents a barrier to using the wake as a DM laboratory.

\citetalias{garavito-camargo_hunting_2019} used a tailored suite of high-resolution \textit{N}-body simulations of the MW/LMC interaction to show that the LMC creates three major responses in the MW's DM halo, work that was later expanded upon by \citet{garavito-camargo_quantifying_2021}. These responses are: 1) the collective response, a large-scale overdensity that leads the LMC and arises primarily due to the shift of the inner halo relative to the outer halo; 2) the global underdensity, which surrounds the LMC’s DF wake; and 3) the DF wake itself.

By ``painting'' a stellar halo onto the DM particles using weighted sampling, \citetalias{garavito-camargo_hunting_2019} also explored the response of the stellar halo to the perturbations induced by the LMC. They found that there should be an observable stellar overdensity associated with the DM wake, which has been tentatively detected \citep{belokurov_pisces_2019, conroy_all-sky_2021}. Further, they found that the velocities of stars in the wake converge near the LMC and diverge behind it, which leads to an enhancement in the component of the stellar velocity dispersion that is orthogonal to the wake. 

While this approach is effective at capturing the global behavior of the MW's DM halo in response to the LMC, it is unable to separate the effect of the DM wake from that of the LMC itself and other halo perturbations. In particular, even in the absence of a DM halo, the passage of a massive perturber such as the LMC would be sufficient to form a wake in the stellar halo \citep{chandrasekhar_dynamical_1943}. If the LMC's wake is to be used as a DM laboratory, a more detailed understanding of the role of the DM wake's self-gravity in forming the stellar wake is required. 

A complementary study by \citet{rozier_constraining_2022} used a linear response formalism to study the effect of the LMC on the MW's dark and stellar halos. An advantage of linear response theory is that it permits disabling the self-gravity of the DM, giving insight into the DM wake's role in shaping the response of the stars. \citet{rozier_constraining_2022} reported that the DM wake's self-gravity enhanced the density of the DM wake by $\sim 10\%$, which hints that the stellar response to the wake is likely sensitive to the density field of the DM wake. This further suggests that the stellar response may also reflect changes in the wake structure owing to the nature of the DM particle.

In particular, the behavior of fuzzy DM (FDM) DF wakes can vary significantly from those in CDM \citep[e.g.][]{bar-or_relaxation_2019, lancaster_dynamical_2020, chavanis_landau_2021, traykova_dynamical_2021, buehler_dynamical_2023, vitsos_dynamical_2023}. FDM is an ultralight bosonic scalar field DM with particle masses of $m_a \sim 10^{-22}$ eV (\citealt{hu_fuzzy_2000}; see also \citealt{hui_ultralight_2017}, \citealt{ferreira_ultra-light_2021}, and \citealt{hui_wave_2021} for reviews), with typical particle de Broglie wavelengths on the order of kpc. FDM exhibits characteristic density fluctuations on size scales comparable to the de Broglie wavelength of the particles, often called ``granules,'' which arise due to wave interference between the particles. In the context of DF, \citet{lancaster_dynamical_2020} and \citet{vitsos_dynamical_2023} showed that FDM granules interact with the perturbing object to produce highly stochastic density fields in the wake, which can result in an oscillatory drag force if the perturber is moving slowly. To test these predictions using the LMC's DF wake, we must first understand whether such an FDM wake would affect the motions and distribution of halo tracers differently than a CDM wake. 

In this paper, we present a suite of windtunnel-style \textit{N}-body simulations of the LMC's DF wake under three different assumptions for the DM model: CDM with self-gravity, CDM without self-gravity, and FDM with self-gravity. We aim to determine the extent to which self-gravity and the assumption of the DM model impact the structure and kinematics of the LMC's DM wake. Additionally, to quantify the effect of the DM wake on the distribution and velocities of halo tracers (halo stars, globular clusters, or satellite galaxies), 
we include a separate population of stellar tracer particles. 

This paper is organized as follows: In $\S$ \ref{sec:sims}, we outline the setup of our windtunnel simulations, including the motivation for our setup, how we choose our initial conditions, and the specifics of each DM model we consider. In $\S$ \ref{sec:DM_results}, we present our results for the structure and kinematics of the DM wakes. $\S$ \ref{sec:star_wakes} discusses the response of the stellar halo to both the LMC and the DM wakes. In $\S$ \ref{sec:discussion}, we introduce a toy model for how the stellar wake might be observed from Earth, and determine the robustness of our results to observational errors. We also explore the effect of the chosen DM model on the LMC's orbit, and discuss the wake's influence as a perturbation to the MW's DM halo. $\S$ \ref{sec:caveats} examines the consequences of changing major assumptions in our simulation setup. Finally, we summarize our findings in $\S$ \ref{sec:conclusion}.

\section{Simulations} \label{sec:sims}

Here, we describe the simulations we use to study the formation of the LMC's DF wake and corresponding response of the MW's stellar halo. In $\S$ \ref{subsec:windtunnel}, we explain the motivation for and the design of our windtunnel setup. $\S$ \ref{subsec:wind_params} and \ref{subsec:stellar_wind_params} describe the motivation for our choices of initial conditions for the DM and stars, respectively. In $\S$ \ref{subsec:CDM_sim}, we describe our CDM simulations which we perform with the \texttt{Gadget4} code \citep{springel_simulating_2021}. Our FDM simulations are performed with the \texttt{BECDM} module \citep{mocz_galaxy_2017, mocz_galaxy_2020} for the \texttt{Arepo} code \citep{springel_e_2010}, and are described in $\S$ \ref{subsec:FDM_sim}.

\subsection{Dark Matter Windtunnels} \label{subsec:windtunnel}

To study the formation of the DM wake behind an LMC-like perturber, we use windtunnel-style simulations, in which the perturber is stationary while a ``wind'' of particles moves past the perturber with a common bulk velocity. The box's boundary conditions are set up such that one boundary acts as an inflow, the opposite boundary acts as an outflow, and the boundaries parallel to the wind's motion are periodic. In this way, the interaction of the perturber with the background wind can be studied in a maximally controlled environment.

Windtunnel setups are commonly used in hydrodynamic simulations (e.g. \citealt{salem_ram_2015}; \citealt{scannapieco_launching_2015}; \citealt{schneider_hydrodynamical_2017}; \citealt{sparre_physics_2019}; \citealt{sparre_interaction_2020}) and when studying DF in FDM backgrounds (\citealt{lancaster_dynamical_2020}; \citealt{vitsos_dynamical_2023}). In hydrodynamic windtunnels, it is common to use inflow/outflow boundaries, where the wind particles are created at the inflow and removed at the outflow. Such boundaries also allow one to change the wind properties with time to mimic a perturber falling deeper into a host galaxy's halo \citep{salem_ram_2015}. In principle, these time-dependent wind properties would seem ideal for our simulations, but in practice, increasing the wind density and speed with time results in the gravitational collapse of the most dense regions of the wind, creating artificial shockwave-like structures.

This restricts us to using a completely uniform background wind, in which the density, dispersion, and velocity remain constant throughout the simulation. Such a wind is most efficiently created by using fully periodic boundary conditions as in \citet{lancaster_dynamical_2020} and \citet{vitsos_dynamical_2023}. When the wind is given a bulk velocity, it loops through the box and naturally creates inflow/outflow-like boundaries. Of course, care must be taken to stop the simulation prior to the wake wrapping through the box boundary, and so all of our simulations are run for one-half the box crossing time at the bulk wind speed. All of our boxes are cubic and have side lengths of $L = 600$ kpc, which allows us to simulate wakes longer than the MW's virial radius.

The LMC in our simulations is represented by an external, stationary Hernquist potential \citep{hernquist_analytical_1990} at the center of the simulation volume. This potential is modeled using the density profile of \citetalias{garavito-camargo_hunting_2019}'s LMC3 (see Table \ref{tab:gal_models}). A uniform background of DM (the DM ``wind'') with constant mass density $\bar{\rho}$ and isotropic velocity dispersion $\bar{\sigma}$ moves across the LMC potential with a constant bulk velocity $v$ in the +$y$-direction. In this study we choose two sets of wind properties, described in Section \ref{subsec:wind_params}. The advantages of this setup over simulating the full LMC-MW interaction with live halos are threefold: 
\begin{enumerate}
    \item For FDM in particular, a windtunnel is \textit{far} less computationally expensive. Specifically, live FDM halos require exceptionally high spatial and temporal resolution (see Section \ref{subsec:FDM_sim}) that makes an FDM simulation analogous to those in \citetalias{garavito-camargo_hunting_2019} prohibitively expensive. A windtunnel, by contrast, requires only that we resolve the relatively uniform wind instead of the complex structure of a halo. 
    \item A windtunnel setup allows us to study the role of the wake's self gravity by running simulations both with and without gravity between DM particles. \textit{N}-body simulations with live halos by nature require self-gravity between their DM particles to keep the halos bound, while a uniform DM wind is not subject to this restriction. This allows us to separate how the MW's stellar halo reacts to \textit{only} the LMC, vs. the LMC plus a DM wake. If this difference is observed, it will provide independent evidence that the LMC is moving through a DM medium. 
    \item Idealized windtunnels present the best stage for studying DF wakes in the absence of other complicating factors present in live interaction simulations such as tides from the host galaxy, the host's reflex motion, and orbital resonances. Our setup thus streamlines analysis because we do not have to disentangle DF from any other process.  
\end{enumerate}

Naturally, the drawback of the windtunnel is that there is no MW potential. As a result, the LMC ``moves'' in a straight line (as opposed to a curved orbit), and the wind speed and density are constant (as opposed to varying as the LMC plunges deeper into the MW's halo). Nevertheless, we use \citetalias{garavito-camargo_hunting_2019}'s fiducial Simulation 3 (their LMC3 and MW1 galaxy models, summarized in Table \ref{tab:gal_models}) as a reference simulation to guide the setup of our simulations in an effort to make our wakes as realistic as possible. 
In Appendix \ref{apdx:live_comp}, we show that the wake in our Fiducial CDM windtunnel simulation closely resembles the wake formed in \citetalias{garavito-camargo_hunting_2019}'s Simulation 3. 

\begin{figure*}
    \centering
    \includegraphics[width=0.41\textwidth]{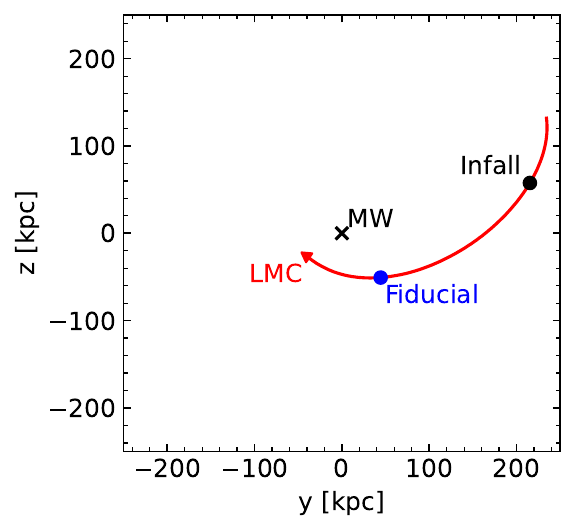}
    \includegraphics[width=0.58\textwidth]{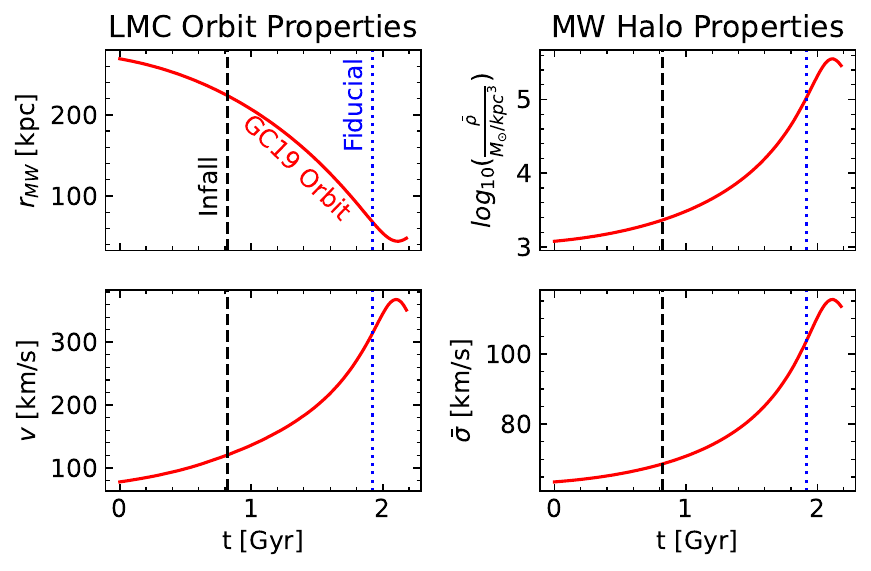}
    \caption{Schematic illustrating how we choose DM windtunnel parameters from the LMC's orbit. The left panel shows the LMC's orbit up to the present day from the reference simulation (simulation 3 in \citetalias{garavito-camargo_hunting_2019}). The orbit (red line) is nearly polar, and it is shown projected onto the $yz$-plane in the rest frame of the MW's center of mass (denoted by the black x). The blue and black points show the locations on the orbit that we use to select the wind parameters for the Fiducial and Infall orbit cases, respectively. The center column of panels shows the LMC's distance from the MW center ($r_{MW}$; top) and orbital speed ($v$; bottom) as a function of time during the reference simulation, while the right column shows the MW's DM halo density ($\bar{\rho}$; top), and velocity dispersion ($\bar{\sigma}$; bottom) at the LMC's location vs. time. Blue dotted and black dashed lines mark where we select the Fiducial and Infall wind parameters, respectively. }
    \label{fig:orbit_cases}
\end{figure*}

\begin{table}[]
    \caption{Summary of the galaxy halo density profiles we use in our simulations. Both profiles are taken from \citetalias{garavito-camargo_hunting_2019}'s simulation 3, and are Hernquist profiles \citep{hernquist_analytical_1990}. We list: our galaxy model; the corresponding galaxy model in \citetalias{garavito-camargo_hunting_2019}; $M$, the total mass of the profile (if it were integrated to infinity); and $a$, the scale radius. For more details on these galaxy models, we refer the reader to \citetalias{garavito-camargo_hunting_2019}. Note that unlike in \citetalias{garavito-camargo_hunting_2019}, neither halo is represented as a live system of \textit{N}-body particles in our simulations, rather the LMC profile is used to set the external potential over which the wind flows, and the MW profile is used to set wind properties as explained in Section \ref{subsec:wind_params}. } \label{tab:gal_models}
    \centering
    \begin{tabular}{c c c c}
        \hline
        \hline
         Galaxy Model & \citetalias{garavito-camargo_hunting_2019} Model & $M$ [M$_\odot$] & $a$ [kpc] \\
         \hline
         MW & MW1 & $1.2 \times 10^{12}$ & 40.1 \\
         LMC & LMC3 & $1.8 \times 10^{11}$ & 20.0 \\
         \hline
         \hline
    \end{tabular}
\end{table}

\subsection{Dark Matter Wind Parameters}\label{subsec:wind_params}

To select the DM wind parameters $\bar{\rho}$, $\bar{\sigma}$, and $v$, we choose a point along the LMC's orbit from our reference simulation, and obtain the Galactocentric position and velocity of the LMC at this point. Then, we calculate $\bar{\rho}$ and $\sigma$ analytically at the orbital radius of interest, using the MW1 density profile from \citetalias{garavito-camargo_hunting_2019} (see Table \ref{tab:gal_models}). The wind bulk velocity $v$ is then simply the LMC's orbital speed. 

Using this procedure to determine wind parameters, we simulate two different cases along the LMC's orbit: 
\begin{itemize}
    \item \citetalias{garavito-camargo_hunting_2019} determine the stellar halo's response to the wake is most easily observed at a Galactocentric distance of 70 kpc to maximize the stellar density while avoiding contamination from the Clouds and the Sagittarius stream. To best reproduce this response with a windtunnel, we want our `Fiducial' CDM wake to match \citetalias{garavito-camargo_hunting_2019}'s wake at 70 kpc, which requires taking the wind parameters from 70 kpc as opposed to the LMC's present-day location or pericenter passage. Therefore, our Fiducial orbit case represents the MW's halo at 70 kpc, when the LMC is moving at $\sim 313$ km/s. 
    \item To study the behavior of FDM vs. CDM wakes and the effect of self-gravity as a function of the LMC's speed and the MW halo's density, we also simulate an `Infall' orbit case. This Infall case represents the MW's halo at a distance of $\sim 223$ kpc (between our MW model's $R_{200}$ and $R_{vir}$), when the LMC is moving at 120 km/s.
\end{itemize}

Figure \ref{fig:orbit_cases} illustrates the selection of these parameters.
In the left panel, we show the LMC's orbit since it first crossed the MW's virial radius, until the present day in the reference simulation. The orbit is projected onto the $yz$-plane, and we mark the locations from which we take each set of wind parameters. Meanwhile, the other panels show the LMC/MW separation, LMC orbital speed, and MW DM density and dispersion at the LMC's location as a function of time. We also mark each choice of windtunnel parameters in each panel. 

For both orbit cases, we run two CDM simulations and one FDM simulation, described in \ref{subsec:CDM_sim} and \ref{subsec:FDM_sim} respectively. See Table \ref{tab:dm_wind} for a summary of the DM wind parameters in each simulation.

\begin{table*}[t]
    \centering
    \caption{Summary of our simulation suite, listing the DM wind properties for each simulation. We list: The orbit case; the dark matter model with the FDM particle mass if applicable; whether DM self gravity is enabled; $r_{\text{MW}}$, the separation between the MW and LMC at the point in the \citetalias{garavito-camargo_hunting_2019} reference simulation that the wind parameters are drawn from; $v$, the wind bulk velocity; the number of DM resolution elements (\textit{N}-body particles for CDM, grid cells for FDM); $\bar{\rho}$, the DM wind density; and $\bar{\sigma}$, the DM wind velocity dispersion. All simulations have a box side length of $L = 600$ kpc. The LMC is represented as a static Hernquist potential and is static at the center of the box throughout each simulation.} \label{tab:dm_wind}
    \begin{tabular}{c|c c c c c c c c c }
        \hline
        \hline
        Orbit Case & DM Model & DM Self-Gravity & $r_{\text{MW}}$ & $v$ & Particles or Cells & $\bar{\rho}$ & $\bar{\sigma}$ \\
        \ & & & [kpc] & [km/s] & & [M$_\odot$/kpc$^3$] & [km/s] \\
        \hline
        \ & CDM & No & 70 & 313.6 & $10^8$ & $1.083 \times 10^5$ & 103.9 \\
        Fiducial & CDM & Yes & 70 & 313.6 & $10^8$ & $1.083 \times 10^5$ & 103.9 \\
        \ & FDM, $m_{a}=10^{-23}$ eV & Yes & 70 & 313.6 & $1024^3$ & $1.083 \times 10^5$ & 103.9 \\
        \hline
        \ & CDM & No & 223 & 120.5 & $10^8$ & $2.315 \times 10^3$ & 68.68 \\
        Infall & CDM & Yes & 223 & 120.5 & $10^8$ & $2.315 \times 10^3$ & 68.68 \\
        \ & FDM, $m_{a}=10^{-23}$ eV & Yes & 223 & 120.5 & $1024^3$ & $2.315 \times 10^3$ & 68.68 \\
        \hline
        \hline
    \end{tabular}
\end{table*}

\subsection{Stellar Wind Parameters} \label{subsec:stellar_wind_params}
In addition to the DM, all of our simulations include a uniform wind of star particles to test the response of the stellar halo to both the LMC and the DM wake. In all simulations, regardless of orbit case or DM model, the stellar wind is composed of test particles at 1 $\rm{M}_\odot$ resolution. Their density is calculated from the K-giant stellar halo density profile of \citet{xue_radial_2015} at a Galactocentric distance of 70 kpc, assuming the stellar halo has a total mass of $10^9 \rm{M}_\odot$ inside the MW's virial radius and is composed entirely of K-giants. The stars' velocity dispersion is 90 km/s, again motivated by measurements at 70 kpc (\citealt{deason_cold_2012}; \citealt{cohen_outer_2017}; \citealt{bird_constraints_2021}), and they move at the same bulk speed as the DM wind. See Table \ref{tab:star_wind_prop} for a summary of the stellar wind properties. We reiterate that while the DM and stellar winds of the Fiducial case are both calibrated for a Galactocentric distance of 70 kpc, we use the same stellar wind for the Infall case (at 223 kpc) as there are few observational constraints on the stellar halo's properties at large distances. 

\begin{table}[]
    \centering
    \caption{Summary of the stellar wind parameters, which are identical in every simulation. We list: $N_\star$, the number of star test particles; $\bar{\rho}_\star$, the stellar wind density; and $\bar{\sigma}_\star$, the stellar wind velocity dispersion. The stellar wind bulk velocity is matched to that of the DM wind in each simulation.} \label{tab:star_wind_prop}
    \begin{tabular}{c|c}
        \hline
        \hline
         Quantity & Value \\
         \hline
         $N_\star$ & $1.257 \times 10^6$ \\
         $\bar{\rho}_\star$ [$M_\odot$/kpc$^3$] & $5.818 \times 10^{-3}$  \\
         $\bar{\sigma}_\star$ [km/s] & 90.00 \\
        \hline
        \hline
    \end{tabular}
\end{table}

\subsection{CDM} \label{subsec:CDM_sim}

Our CDM simulations are performed with the \texttt{Gadget-4} \textit{N}-body and smoothed particle hydrodynamics code \citep{springel_simulating_2021}. We use $10^8$ DM particles in all CDM simulations, which results in a mass resolution of $5.0 \times 10^3 \rm{M}_\odot$ for the infall wind, and $2.3 \times 10^5 \rm{M}_\odot$ for the fiducial wind. All simulations use a softening length of 0.16 kpc, from Equation 15 of \citet{power_inner_2003} with \citetalias{garavito-camargo_hunting_2019}'s MW1 model. 

For our CDM initial conditions, we begin by determining the particle mass based on the box volume, number of particles, and the desired wind density $\bar{\rho}$. Particle positions are set randomly throughout the box to create a wind of uniform density. All three velocity components are sampled from a Gaussian according to the isotropic velocity dispersion $\bar{\sigma}$. Finally, every particle is boosted by the bulk wind velocity $v$ in the +$y$-direction. 

An identical procedure is used to create the star initial conditions for all simulations in our suite, though we note again that the stellar wind uses a different density and velocity dispersion than the dark matter wind (see Table \ref{tab:star_wind_prop}).

For each orbit case, we run two CDM simulations: one without self-gravity between the DM particles (i.e. the ONLY forces on simulation particles are from the LMC), and one with self-gravity between the DM particles but NOT the star particles (i.e. all particles feel gravity from the LMC and DM particles, but not from the stars). Comparisons between these simulations allow us to isolate the effects of the DM wake's self-gravity from the influence of the LMC.

\subsection{FDM} \label{subsec:FDM_sim}

Our FDM simulations are performed using the \texttt{BECDM} module (\citealt{mocz_galaxy_2017}; \citealt{mocz_galaxy_2020}) for the \texttt{Arepo} code \citep{springel_e_2010}. \texttt{BECDM} uses a second-order pseudo-spectral method to solve the FDM equations of motion on a discretized, fixed grid, similar to the \texttt{AxiREPO} module introduced by \citet{may_structure_2021} and \citet{may_halo_2023}. 

For more detailed background on FDM as a DM candidate, we refer the reader to reviews by \citet{hui_ultralight_2017}, \citet{ferreira_ultra-light_2021}, \citet{hui_wave_2021}, and references therein. For detailed descriptions of the numerical methods used here, we refer the reader to \citet{mocz_galaxy_2017}, \citet{mocz_galaxy_2020}, \citet{may_structure_2021}, \citet{may_halo_2023}, and references therein. However, we provide an abridged description and information specific to our windtunnel simulations here for completeness. 

The FDM is described by a single wavefunction, which takes the form of a complex-valued scalar field 

\begin{equation} 
    \psi = \sqrt{\rho}e^{i\theta} ,
    \label{eqn:wavefunc}
\end{equation}

\noindent where $\rho = |\psi|^2$ is the mass density of the FDM and $\theta \in [0,2\pi)$ is the phase. $\psi$ obeys the Schr\"odinger-Poisson (SP) equations of motion in the non-relativistic limit:

\begin{equation}
    i \hbar \frac{\partial \psi}{\partial t} = \left[ - \frac{\hbar^2}{2m_a} \nabla^2 + m_a V \right] \psi
    \label{eqn:schro}
\end{equation}

\begin{equation}
    \nabla^2 V = 4 \pi G (\rho - \bar{\rho}) ,
    \label{eqn:poiss}
\end{equation}

\noindent where $m_a$ is the FDM particle mass, and $V$ is the gravitational potential. Additionally, the velocity field of the FDM is encoded by the phase $\theta$ via 

\begin{equation}
    \vec{u} = \frac{\hbar}{m_a}\nabla \theta ,
    \label{eqn:phase}
\end{equation}

\noindent where $\vec{u}$ is the velocity of the FDM. 

The FDM wavefunction (Equation \ref{eqn:wavefunc}) is discretized onto a grid of $N^3$ cells of size $dx = L/N$, where $L$ is the side length of the simulation box, and evolved using a kick-drift-kick algorithm. During one timestep $dt$, the potential is first calculated as 

\begin{equation}
    V = \text{ifft}(-\text{fft} (4 \pi G (\rho - \bar{\rho})/k^2) + V_\text{LMC},
    \label{eqn:potential}
\end{equation}

\noindent where fft and ifft indicate fast-Fourier and inverse fast-Fourier transforms, respectively, $k$ is the wavenumbers associated with the grid cells, and $V_\text{LMC}$ is the external LMC potential. Then, the first ``kick'' is performed using half the timestep:

\begin{equation}
    \psi \leftarrow \text{exp}\left[-i\frac{m_a}{\hbar}\frac{dt}{2}V\right]\psi
    \label{eqn:FDMkick}
\end{equation}

\noindent Next is the ``drift,'' performed in Fourier space as

\begin{equation}
    \hat{\psi} = \text{fft} (\psi)
\end{equation}
\begin{equation}
    \hat{\psi} \leftarrow \text{exp} \left[-i \frac{\hbar k^2}{2 m_a} dt\right] \hat{\psi}
    \label{eqn:FDMdrift}
\end{equation}
\begin{equation}
    \psi = \text{ifft} (\hat{\psi}) 
\end{equation}

\noindent and finally, the timestep is completed by an additional half-step ``kick'' via Equation \ref{eqn:FDMkick}. 

Directly solving the SP equations as we do here has the advantage that it self-consistently describes the full wave dynamics of the FDM, including interference patterns (sometimes called ``granules'' or ``fringes'') that arise from the velocity dispersion of the FDM and interactions with the LMC potential. Capturing the full wave behavior of the FDM is especially important in studies of DF, as the interference patterns that arise in FDM DF wakes can cause significant deviations from CDM, including stochastic oscillation of the drag force \citep{bar-or_relaxation_2019, lancaster_dynamical_2020, traykova_dynamical_2021, buehler_dynamical_2023, vitsos_dynamical_2023}. Other numerical descriptions of FDM such as SPH methods or fluid dynamics approaches via the Madelung transformation \citep{madelung_quantentheorie_1927} either approximate or ignore the detailed wave behavior. 

The disadvantage of directly solving the SP equations is the enormous spatial and temporal resolution required for numerical convergence. The resolution criteria arise from the wavefunction phase $\theta$, which cannot vary by more than $2\pi$ in a grid cell during one timestep (which gives the temporal resolution requirement), or between adjacent grid cells in the same timestep (which gives the spatial resolution requirement). 

To satisfy the temporal resolution requirement, \texttt{BECDM} uses the timestep criterion

\begin{equation}
    dt \leq \text{max} \left[\frac{m_a}{6\hbar} dx^2, \frac{h}{m_a |V|_\text{max}}\right]
    \label{eqn:fdm_dt}
\end{equation}
where $|V|_\text{max}$ is the maximum of the absolute value of the potential (\citealt{schwabe_simulations_2016}; \citealt{mocz_galaxy_2017}).  

The spatial condition may equivalently be thought of as the requirement that all velocities are resolved, i.e. that the largest velocity in the simulation does not exceed $2\pi\hbar/m_a dx$ (see Equation \ref{eqn:phase}), or that the smallest de Broglie wavelengths in the problem are resolved:

\begin{equation}
    dx = \frac{L}{N} \leq \frac{h}{m_au_\text{max}}
\end{equation}

In practice, to ensure that the largest velocities in our simulations are well below $u_{\rm{max}}$, we set the limit on $dx$ according to the bulk wind velocity (the largest velocity scale in the simulation) and then divide by a further factor of $2\pi$, such that our grid cell sizes follow

\begin{equation} \label{eqn:grid_size}
    dx \leq \frac{\hbar}{m_av}.
\end{equation}

\noindent For $m_a = 10^{-23}$ eV and our highest (Fiducial) wind speed of 313.6 km/s, the right-hand side evaluates to 0.611 kpc, which satisfies Equation \ref{eqn:grid_size} when $dx = 600$ kpc$/1024=0.586$ kpc.

To generate our FDM initial conditions, we take advantage of the property that $\psi$ can be constructed according to a desired distribution function $f$ as

\begin{equation}
    \psi(\vec{x}) \propto \sum_{j=0}^{N^3} \sqrt{f(\vec{x}, \vec{u}_j)} \text{exp} \left[ i \frac{m_a}{\hbar} \vec{x}\cdot\vec{u}_j + i2\pi (\phi_{\text{rand},j}) \right] ,
    \label{eqn:psi_df}
\end{equation}

\noindent where the sum is over all grid cells in 3-D, and $\phi_{\text{rand},j} \in [0, 2\pi)$ is a random number that ensures the phases of each mode are random and uncorrelated, i.e. the FDM has some isotropic velocity dispersion \citep{widrow_using_1993}. In practice, we desire an FDM wind that is equivalent to our CDM wind, such that it is uniform on the scale of the box and follows an isotropic, Maxwellian velocity distribution. To do this, we take the equivalent approach of constructing the initial conditions in frequency space before taking the inverse Fourier transform and then normalizing such that the mean FDM density is the desired wind density $\bar{\rho}$:

\begin{equation} 
    \hat{\psi} \propto \sqrt{\text{exp} \left[ - \left( \frac{\hbar}{m_a}\right)^2 \frac{(2\pi/L)^2 k^2}{2\sigma^2} \right]} e^{i 2 \pi \phi_{\text{rand},j}}
    \label{eqn:fdm_distribution_func}
\end{equation}

\begin{equation}
    \psi = \text{ifft} (\hat{\psi}) 
\end{equation}

\begin{equation}
    \psi \leftarrow \psi \sqrt{\bar{\rho} / \left(\frac{1}{N^3}\sum_{j=0}^{N^3}|\psi_j^2|\right)}
\end{equation}

Finally, we apply the bulk wind velocity boost by calculating the wavenumber associated with the desired wind velocity

\begin{equation}
    k_{\text{boost}} = \frac{vm_aL}{2\pi\hbar}
\end{equation}

\noindent and then applying the boost via

\begin{equation}
    \psi \leftarrow \psi \ \text{exp} \left[ i k_{\text{boost}} \frac{2 \pi y}{L} \right].
\end{equation}

For each orbit case (Fiducal, Infall; see Table \ref{tab:dm_wind} and Figure \ref{fig:orbit_cases}), our primary choice for the FDM particle mass is $10^{-23}$ eV. This is the largest particle mass that is feasible to simulate with $N = 1024$ and $L = 600$ kpc.\footnote{Our FDM simulations each take $\sim 370000$ CPU hours at this resolution. We are restricted to $L \geq 600$ kpc to simulate a sufficiently long wake, so increasing the particle mass by a factor of just two requires $2048^3$ cells. Using the characteristic $\mathcal{O}(N \rm{log}(N))$ scaling of the FFT calculations that BECDM relies on, such simulations would take $\sim 800000$ CPU hours each, which we consider prohibitively expensive.} 

Lastly, we justify our choice to use the same wind parameters for both our FDM and CDM simulations, as FDM halos differ fundamentally from CDM halos. Instead of being constructed from individual DM particles that obey a particular distribution function (as in CDM), FDM halos are better described as a superposition of eigenmodes that combine to produce a ground-state soliton core surrounded by a ``skirt'' of excited states that follow an NFW-like \citep{navarro_universal_1997} density profile \citep[e.g.][]{schive_cosmic_2014, mocz_galaxy_2017, may_structure_2021, chan_diversity_2022, yavetz_construction_2022, zagorac_schrodinger-poisson_2022}. Thus, it is important to verify that our choice of DM wind parameters $v$, $\bar{\rho}$, and $\bar{\sigma}$ is reasonable in FDM given that we motivate them from a CDM simulation.

As described in $\S$ \ref{subsec:wind_params}, $v$ is given by the LMC's orbital speed, while $\bar{\rho}$ and $\bar{\sigma}$ come from the MW's halo. We discuss each parameter in turn:

\begin{itemize}
    \item In $\S$ \ref{subsec:drag_forces}, we argue that the LMC's orbit is the same in both a CDM and FDM universe, so our choices of $v$ are valid in both DM models.
    \item The MW halo's density profile is expected to match in CDM and FDM provided we are interested in a regime well outside the soliton core such that the FDM halo follows an NFW-like density profile similar to a CDM halo. \citet{schive_cosmic_2014} show that the MW's soliton would have a radius of $\approx$ 0.18 kpc, so at the orbital distances of the LMC ($\geq 40$ kpc) we expect our choices of $\bar{\rho}$ to be valid in both DM models.
    \item \citet{yavetz_construction_2022} show in their Appendix A that far from the soliton core, there is a direct correspondence between the classical particle distribution function of a CDM halo and the eigenmodes that comprise an FDM halo. As such, we expect that for the region of interest in our windtunnel (i.e. a volume many times larger than the de Broglie wavelength and far from the core), using a CDM distribution function to set the FDM eigenmodes (Equation \ref{eqn:fdm_distribution_func}) is a reasonable approach (T. Yavetz, personal communication 2023).
\end{itemize}

Ultimately, we expect our choice of wind initial conditions to be equally valid in CDM and FDM. It is also worth noting that the inner density profile of the LMC would likely be different in FDM due to the presence of a core. However, in this work we use the same LMC model in both our CDM and FDM simulations to ensure that any differences in our wakes are due purely to our choice of DM model and not the density profile of the perturber. We leave an investigation of the wake's dependence on the perturber's density profile to future work.

\section{Dark Matter Wakes} \label{sec:DM_results}

In this section, we compare the structure and kinematics of the DM wakes in 1) CDM without self-gravity, 2) CDM with self-gravity, and 3) FDM with $m_a = 10^{-23}$ eV. 

\subsection{Density} \label{subsec:DM_density}

\begin{figure*}
    \centering
    \includegraphics[width=1.0\textwidth]{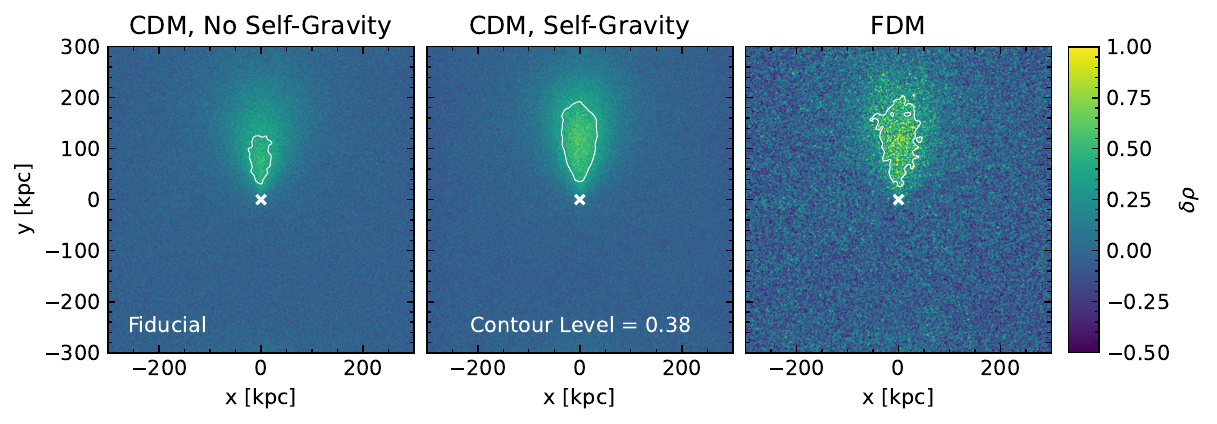}
    \caption{DM overdensity of simulations from the Fiducial case after $t=0.7$ Gyr of integration. Each panel shows the projected overdensity (Equation \ref{eqn:odens}) in a 120 kpc thick slice of the box centered on the LMC potential (i.e. we select particles with $z \in [-60, 60]$) for different DM models. A white cross in each panel denotes the center of the LMC potential. In each panel, the wake manifests as a net overdensity trailing the LMC. The left panel shows the CDM wake without self-gravity, while the center panel shows the CDM wake with self-gravity. To compare the strength and extent of the wake across the panels, the contour in each panel encloses the region with an overdensity greater than the half-max of the CDM wake with self-gravity, which is 0.38. Comparing the two CDM simulations, the addition of self-gravity increases the size and strength (see also Figure \ref{fig:int_dm_density_profile}) of the wake. Specifically, while the wake without self-gravity decays to $\delta \rho < 0.38$ $\sim 130$ kpc behind the LMC, the wake with self-gravity takes $\sim 200$ kpc to do the same. This indicates self-gravity has a significant effect on the wake structure, keeping the wake coherent over larger distances. The right panel shows the FDM wake. The signature ``granular'' structure of the FDM is apparent both in the background and in the wake. Some granules within the wake reach much higher overdensities than either CDM wake achieves, though the FDM wake is still qualitatively similar to the CDM wake with self-gravity. 
    }
    \label{fig:int_dm_density}
\end{figure*}

\begin{figure}
    \centering
    \includegraphics[width=1.0\columnwidth]{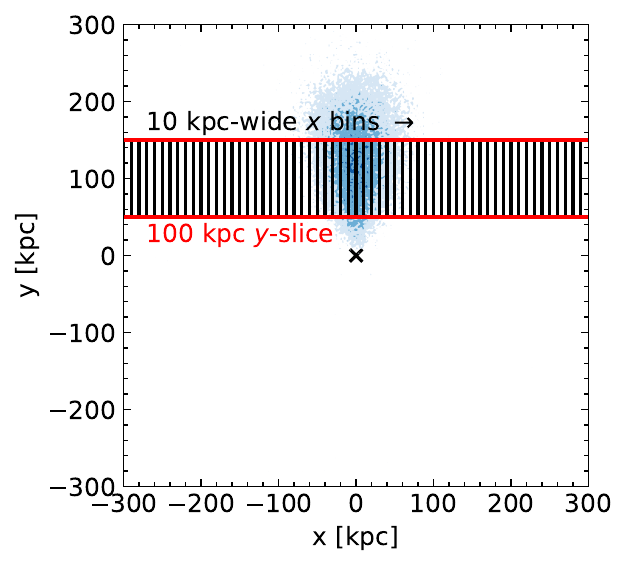}
    \caption{Schematic of the binning procedure to make 
    wake profile plots, in which we compute a quantity of interest across the width of the wake. The filled contours show the typical location of a wake behind the LMC potential, whose center is shown by the x. We begin by selecting particles/cells from  $z \in [-60, 60]$, as in the projection plots (e.g. Figure \ref{fig:int_dm_density}). Then, as pictured, we take a 100-kpc-wide slice of the box of $y \in [50, 150]$ (red), and select particles within 10-kpc-wide bins according to their $x$-position (black). In each $x$-bin, we calculate the quantity of interest, either density (as in Figures \ref{fig:int_dm_density_profile} and \ref{fig:start_dm_density_profile}, or $z$-velocity dispersion (as in Figures \ref{fig:int_dm_radDisp_profile} and \ref{fig:start_dm_radDisp_profile}. For each figure, this process is repeated for five snapshots spanning 100 Myr. The corresponding profile plots illustrate the average of the five snapshots.
    }
    \label{fig:prof}
\end{figure}

\begin{figure}
     \includegraphics[width=1.0\columnwidth]{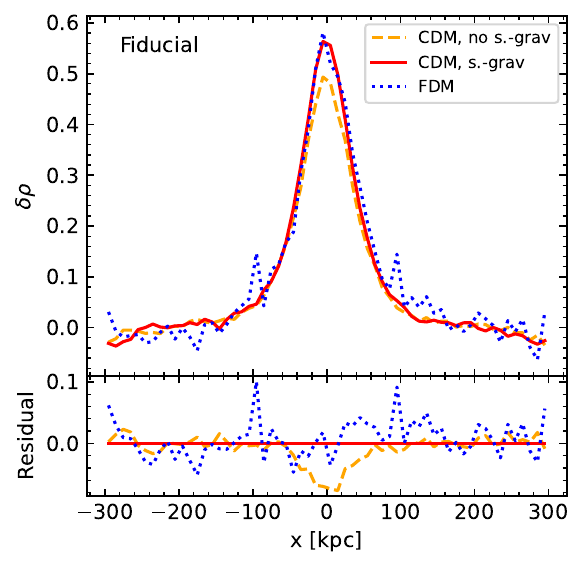}
     \caption{Time-averaged density profile plots for the simulations in Figure \ref{fig:int_dm_density}, showing how the overdensity varies across the wake in the $x$-direction for each DM model over the past 100 Myr.
     The procedure for making these plots is explained in Figure \ref{fig:prof} and in the text. The upper panel shows the overdensity as a function of $x$, while the lower panel shows the residuals of the upper panel with respect to CDM with self-gravity. The CDM simulation without self-gravity is shown as the orange dashed line, the CDM simulation with self-gravity is the red solid line, and the FDM simulation is the blue dotted line. The peak overdensity of the CDM wake without self-gravity is $\sim 10 \%$ lower than in the CDM simulation with self-gravity included, demonstrating the addition of self-gravity increases the peak wake density. The density across the FDM wake oscillates about that of the CDM wake with self-gravity included, with an amplitude of $\delta \rho \sim 0.05$. 
     }
     \label{fig:int_dm_density_profile}
\end{figure}

Figure \ref{fig:int_dm_density} shows the density structure of the simulations with the Fiducial wind (see Table \ref{tab:dm_wind})
for our three primary DM models/scenarios.
In this figure and throughout this work, when we discuss the density of simulation particles, we will use the overdensity 
\begin{equation} \label{eqn:odens}
    \delta \rho = \rho/\bar{\rho} - 1 = \Delta \rho / \bar{\rho}, 
\end{equation}

\noindent which measures the relative change of the density compared to the input wind density, i.e. an overdensity of 0.1 corresponds to a $10\%$ increase in density over the background. Figure \ref{fig:int_dm_density} shows the projected overdensity of each simulation after they have been evolved for 0.7 Gyr,\footnote{The wind travels $\approx 225$ kpc during this time} which is the latest time at which there is no evidence that the wake has begun wrapping through the box's periodic boundaries.

In Figure \ref{fig:int_dm_density}, we begin by taking a 120-kpc wide slice about the box's midplane in $z$, that is we select particles/cells with $z \in [-60, 60]$. For the CDM simulations, we then calculate the projected (column) density of DM particles in a grid of 2 kpc wide bins in $x$ and y, before calculating the overdensity according to Equation \ref{eqn:odens}. For the FDM simulation, we calculate and display the column overdensity in each $z$-column of cells in the 120-kpc slab with the same $x$-$y$  coordinates. The white cross in each panel marks the location of the center of the LMC potential. In each simulation, the DM wake is apparent as an overdensity extending from the center of the box in the $+y$ - direction. To ease comparison between the DM models, we calculate the half-max of the overdensity in the CDM simulation with self-gravity ($\delta \rho = 0.38$), and enclose the region with $\delta \rho$ higher than this with a contour in each panel. When placing the contours, we smooth the density with a Gaussian kernel of $\sigma = 4$ kpc, which reduces the noise associated with the FDM granules. 

The two leftmost panels show the two CDM simulations. Comparing these two panels, the DM wake becomes larger when adding self-gravity: in the left panel (without self-gravity), the region enclosed by the contour reaches a maximum width of $\sim 50$ kpc and extends $\sim 130$ kpc behind the LMC. Adding self-gravity (middle panel) increases the width of the contour to $\sim 80$ kpc, and the length to $\sim 200$ kpc. Importantly, the augmentation in wake length demonstrates that the DM wake's self-gravity plays a significant role in the wake's structure, acting to hold the wake together at larger distances behind the LMC. 

The right panel shows the FDM simulation (with $m_a = 10^{-23}$ eV; see Table \ref{tab:dm_wind}).  We stress that the relative fuzziness of the FDM wakes is \textit{not} a resolution effect (in fact, the FDM simulation is at higher resolution than the CDM). Rather, this granularity is a characteristic property of the FDM that arises due to wave interference between the FDM particles in a velocity-dispersed medium. The FDM wake looks qualitatively similar to the CDM wake with self-gravity aside from the granularity. While some granules near the center of the wake reach much higher overdensities than are seen in CDM, these granules are small and the overall density structure is qualitatively similar to the CDM wake. In $\S$ \ref{subsec:FDM_mass}, we will discuss the impact of FDM particle mass on these results. 

We quantify the wake overdensity by plotting a time-averaged, cross-sectional profile of the wake along the $x$-direction (perpendicular to the wind motion). Figure \ref{fig:prof} describes this process in a schematic. We begin by taking the same $z$-slice as we do for the projection plots ($z \in [-60, 60]$). Then, we select particles/cells in a 100 kpc-thick slice in $y \in [50, 150]$ just behind the LMC, before binning the particles/cells along the $x$-direction in 10 kpc wide bins. Within each $x$-bin, we calculate the overdensity. To reduce noise and limit errors related to our choice of a specific snapshot, we repeat this process for five time-adjacent snapshots, spanning 100 Myr. The density in each $x$-bin is then averaged over the five snapshots, giving us a time-averaged profile of the density as a function of $x$ across the wake. 

Figure \ref{fig:int_dm_density_profile} shows the resulting profiles of the overdensity across the wakes generated in our Fiducial simulations (Figure \ref{fig:int_dm_density})
from $t=0.6-0.7$ Gyr. The upper panel shows the overdensity of each DM wake as a function of $x$, i.e. across the wake, and the lower panel shows the residuals with respect to CDM with self-gravity. The wakes show up as strong density peaks at the center of the box. The addition of self-gravity to the CDM wake raises the peak overdensity by roughly 10\%, in agreement with the results of \citet{rozier_constraining_2022}. The granularity of the FDM wake shows up as oscillations with an amplitude of $\delta \rho \sim 0.05$, though the average profile of the FDM wake matches the CDM wake with self-gravity.

\begin{figure*}
    \centering
    \includegraphics[width=1.0\textwidth]
    {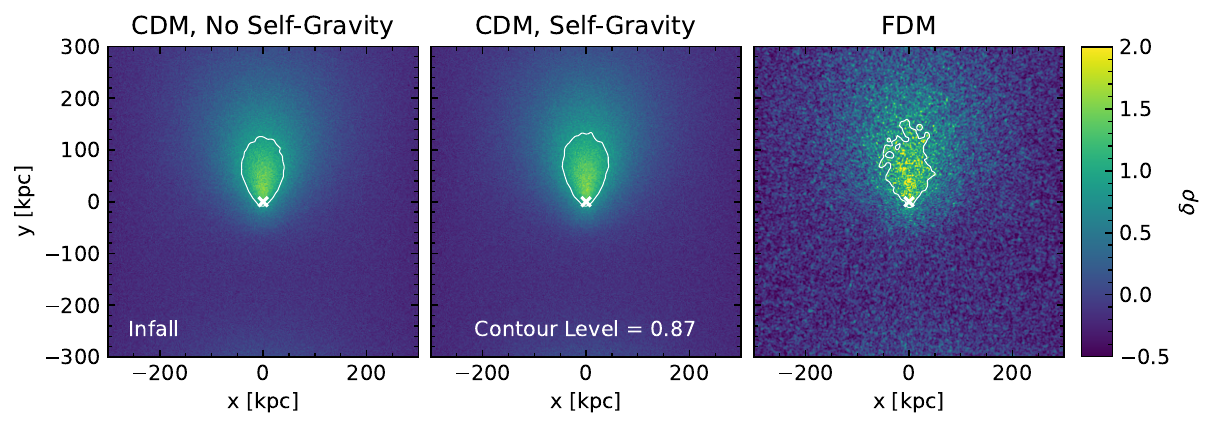}
    \caption{Overdensity of the Infall orbit case simulations after $t=2$ Gyr of evolution (similar to as Figure \ref{fig:int_dm_density}). Comparing the CDM simulations in the left and center reveals that the effect of the wake's self-gravity is almost negligible at this speed. The CDM with self-gravity half-max contour is placed at $\delta \rho = 0.87$, and extends to $\sim 130$ kpc behind the LMC in both CDM simulations. The FDM wake (right panel) is once again more granular than the CDM wakes, though overall it remains qualitatively similar to CDM. At this slower wind speed, the wakes reach higher overdensities and are wider when compared to the Fiducial Case (i.e. the max. width of the contour is $\sim 80$ kpc wide in the Fiducial case, compared to $\sim 100$ kpc in the Infall case here). 
    } 
    \label{fig:start_dm_density}
\end{figure*}

\begin{figure}
    \centering
    \includegraphics[width=1.0\columnwidth]{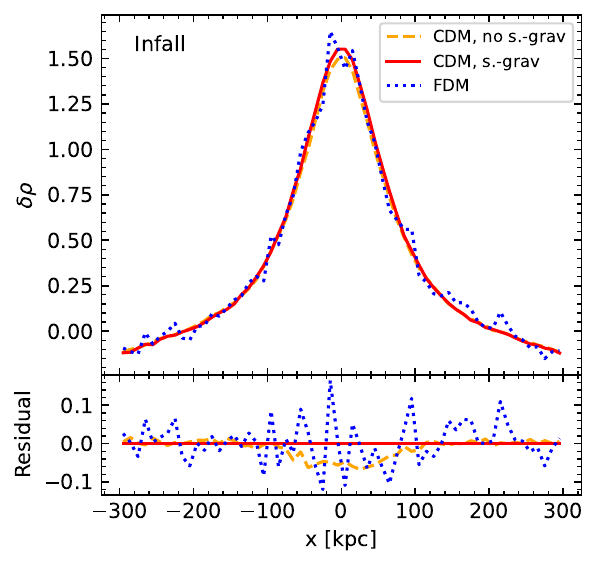}
    \caption{Density profile (same as Figure \ref{fig:int_dm_density_profile}), but for the Infall orbit case after $\sim$2 Gyr of evolution, averaged over 5 snapshots spanning 100 Myr. The CDM wake with self-gravity reaches a peak overdensity of $\sim 1.6$, much higher than in the Fiducial case ($\sim 0.56$; see Figure \ref{fig:int_dm_density_profile}). Removing self-gravity lowers the peak overdensity only by $\sim 3\%$ in the Infall case, showing that the wake's self-gravity is relatively unimportant in this regime. The FDM density continues to oscillate about the CDM density with an amplitude of $\delta \rho \sim 0.05$.} 
    \label{fig:start_dm_density_profile}
\end{figure}

Figure \ref{fig:start_dm_density} is the same as Figure \ref{fig:int_dm_density} but for the Infall orbit case simulations after 2 Gyr (again the last timestep at which there is no evidence for the wake wrapping through the box). As a reminder, the wind in this case is roughly 100 times less dense and moving 1/3 as fast as the wind in the Fiducial case (see Table \ref{tab:dm_wind}). The lower wind speed means particles spend a longer time near the LMC, creating a wider wake when compared to the Fiducial case: in the CDM simulation with self-gravity, the contour is $\sim 20$ kpc wider in the Infall case than the Fiducial case (compare to Figure \ref{fig:int_dm_density}).

The slower speed greatly reduces the effect of the wake's self-gravity, as the relative importance of the LMC's influence on the particles' motions increases.
Comparing the CDM simulations (two leftmost panels) shows they are now almost indistinguishable in projection. Just as in the Fiducial case, the FDM wake appears similar to the CDM wake with self-gravity but is more granular. 

\begin{figure*}
    \centering
    \includegraphics[width=1.0\textwidth]{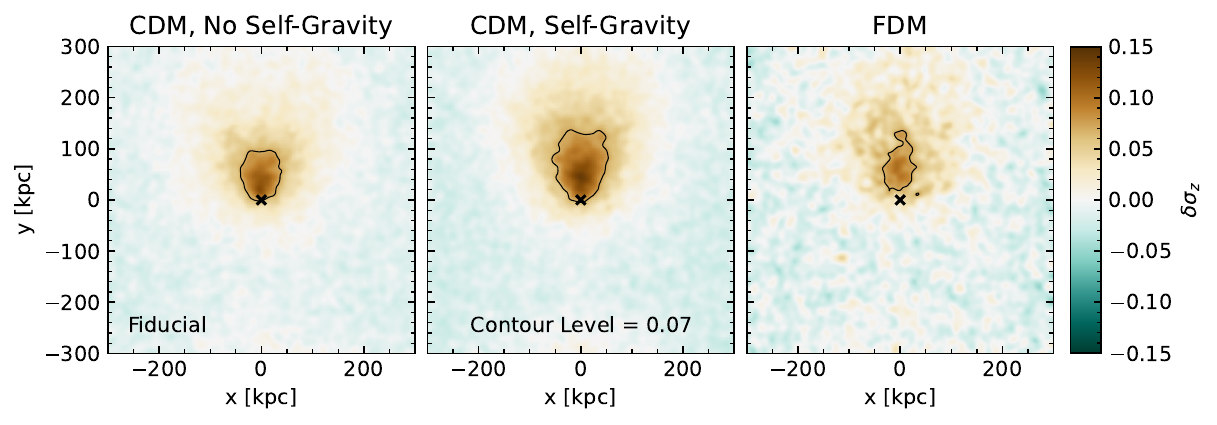}
    \caption{Same as Figure \ref{fig:int_dm_density}, but showing the $z$ (into/out of the page) component of the velocity dispersion for the Fiducial Case. The wake shows up as an enhancement of the velocity dispersion behind the LMC as DM particles converge. The dispersion is computed in a similar fashion to the overdensity: the simulated dispersion is divided by the mean dispersion, then subtract 1 (see Equation \ref{eqn_odisp}). The effect of self-gravity in the CDM simulations is evident, as the region enclosed by the contour is larger ($\sim 20$ kpc wider and $\sim 40$ kpc longer) with self-gravity. The FDM granularity is apparent in the velocity dispersion signature as well. Overall, the granularity results in a weaker dispersion signature in FDM compared to CDM: while the contour in the FDM simulation reaches the same length as in the CDM simulations with self-gravity, it is $\sim 40$ kpc thinner and tapers more quickly than CDM without self-gravity. We discuss this effect further in Figure \ref{fig:int_dm_radDisp_profile} and in the text.}
    \label{fig:int_dm_radDisp}
\end{figure*}

\begin{figure}
    \centering
       \includegraphics[width=1.0\columnwidth]{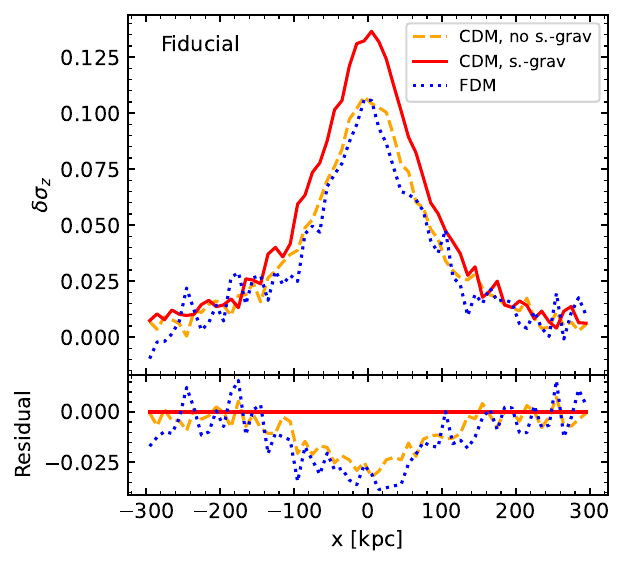}
    \caption{The profile of the $z$-velocity dispersion computed across the wakes in the Fiducial simulations, shown in the same fashion as the overdensity (Figures \ref{fig:int_dm_density_profile} and \ref{fig:start_dm_density_profile}). Results are averaged over 5 snapshots spanning $t=0.6-0.7$ Gyr of evolution. The addition of self-gravity to the CDM simulations raises the peak of the dispersion by $\delta \sigma_z \sim 0.03$. The FDM dispersion profile across the wake is consistently weaker than either CDM case, though the FDM peak is very close to the peak in CDM without self-gravity. Overall, the FDM wake is colder than the CDM wake.}
    \label{fig:int_dm_radDisp_profile}
\end{figure}

The density profiles in Figure \ref{fig:start_dm_density_profile} reinforce this result, as we see the density profiles across the wakes of the two CDM simulations are very close, only showing a $\sim 3\%$ difference at the peak. Meanwhile, the FDM wake's density oscillates about the CDM simulation with self-gravity with an amplitude of $\delta \rho \sim 0.05$ as in the Fiducial wind case. An additional effect of the slower wind speed is that the wakes in the Infall case reach much higher overdensity peaks ($\delta \rho \sim 1.6$, compared to $\sim 0.56$ in the Fiducial case (see Figure \ref{fig:int_dm_density_profile}).

Overall, these results imply that the wake's self-gravity is only expected to become relevant at higher orbital speeds, i.e. once the LMC reaches a Galactocentric distance of $\sim 100$ kpc. Therefore, observable effects of the wake's self-gravity (i.e. halo tracers' reaction to the wake) will likely not be present outside of $\sim 100$ kpc. Meanwhile, the SMC is at a distance of 60 kpc and extends $\sim 30^o$ on the sky from the LMC \citep{grady_magellanic_2021}. The LMC's orbit extends past the SMC on the sky at a distance of $\sim70$ kpc \citepalias{garavito-camargo_hunting_2019}. Together, the decreased effect of DM self-gravity outside of 100 kpc and the need for avoiding SMC contamination suggest the effects of the wake's self-gravity are best searched for at distances of 70-100 kpc. 

\subsection{Velocity Dispersion} \label{subsec:DM_disp}

\begin{figure*}
    \centering
    \includegraphics[width=1.0\textwidth]{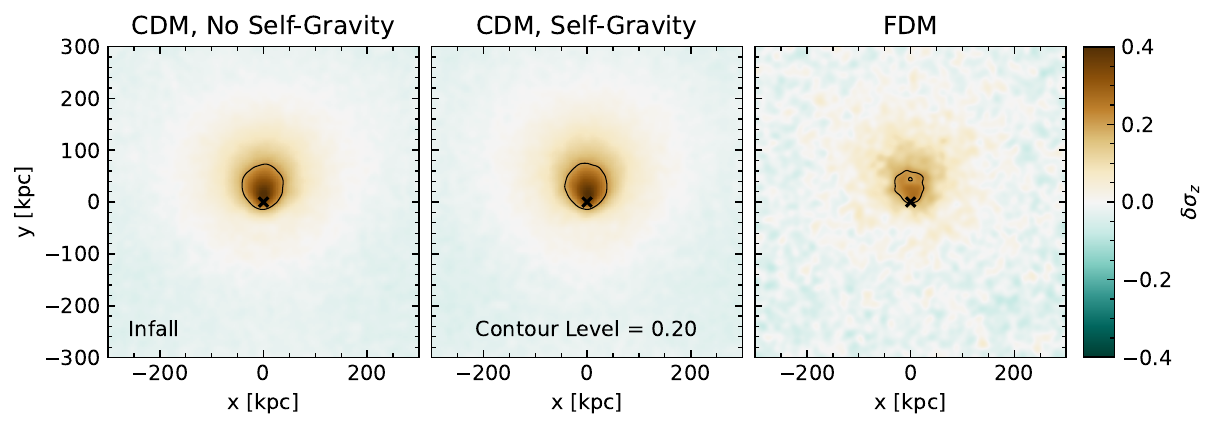}
    \caption{$z$-velocity dispersion for the Infall orbit case simulations, where the simulations are evolved for 2 Gyr. The two CDM simulations agree closely due to the reduced importance of the wake's self-gravity at this lower wind speed and density. The velocity dispersion within the FDM wake is lower than in all CDM wakes, as the region enclosed by the contour is $\sim 30$ kpc shorter and thinner in the FDM case. Overall, the velocity dispersion enhancement is larger in the Infall case than the Fiducial case, as particles spend more time near the LMC, owing to the lower orbital speed (120.5 km/s vs. 313.6 km/s). Consequently, the DM is more deflected by the LMC.}
    \label{fig:start_dm_radDisp}
\end{figure*}

\begin{figure}
    \centering
    \includegraphics[width=1.0\columnwidth]{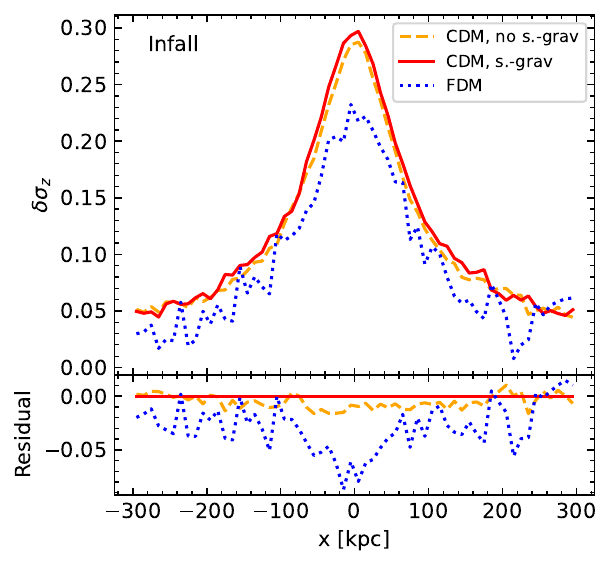}
    \caption{Profile plots of the $z$-velocity dispersion across the wake in the Infall simulations. The CDM simulations agree closely, with the self-gravity-on version having a slightly larger ($\delta \sigma_z \sim 0.01$) peak. The FDM wake is now much colder than either CDM wake, with a peak $\delta \sigma_z \sim 0.07$ lower than either CDM simulation.}
    \label{fig:start_dm_radDisp_profile}
\end{figure}

Figure \ref{fig:int_dm_radDisp} shows the $z$-velocity dispersion of the DM in each simulation of the Fiducial case, analogous to Figure \ref{fig:int_dm_density}. We follow the same binning procedure as in the previous section, with a few small differences: The  $z$-slice is still from $z \in [-60,60]$, however, we use an $x$-$y$ grid of 3 kpc bins, and calculate the $z$-velocity dispersion in each bin. Finally, we apply a 6 kpc-wide Gaussian smoothing kernel. Similar to the overdensity, we report the dispersion as its relative difference from the mean dispersion, which we refer to as the velocity dispersion enhancement: 

\begin{equation} \label{eqn_odisp}
    \delta \sigma_z = \sigma_z / \bar{\sigma} - 1 = \Delta \sigma_z / \bar{\sigma}, 
\end{equation}

\noindent We also include a single contour placed at the half-max of the CDM simulation with self-gravity.

The wake signature is an increase of the dispersion resulting from particles being deflected as they move past the LMC. Comparing the two CDM simulations in Figure \ref{fig:int_dm_radDisp} (left and center panels), the effects of the wake's self-gravity on the velocity dispersion are similar to the density: when self-gravity is turned on, the wake becomes larger. Specifically, the region enclosed by the contour becomes $\sim 40$ kpc longer and $\sim 20$ kpc wider.

For the FDM simulation (right), the granularity is still present in the velocity dispersion, causing an oscillatory behavior that washes out the smooth wake. The contour is much more irregular in shape, and encloses a $\sim 40$ kpc narrower region than in CDM with self-gravity.

In Figure \ref{fig:int_dm_radDisp_profile}, we compute the dispersion profile across the simulated wakes. We calculate these profiles identically to their density versions (Figures \ref{fig:int_dm_density_profile} and \ref{fig:start_dm_density_profile}), where, instead of overdensity, we calculate the $z$-velocity dispersion enhancement in each bin. 

Here, we again see a stronger ($\delta \sigma_z \sim 0.03$) peak in the CDM wake when self-gravity is on versus when it is not included. Interestingly, unlike the density, the mean of the FDM wake's oscillations does \textit{not} trace the CDM wake with self-gravity. Instead, the FDM profile is consistently similar to the CDM profile \textit{without} self-gravity, showing that a self-gravitating FDM wake is colder than a self-gravitating CDM wake.

Figure \ref{fig:start_dm_radDisp} shows the $z$-velocity dispersion within the simulated wakes, but for the Infall orbit case. Overall, we see that the slower wind speed results in a stronger but less extended (in the $y$-direction)
response in velocity dispersion compared to the Fiducial case. 

Like the density, the CDM wakes show a much smaller difference in the Infall case, as the LMC has more time to influence the particle velocities. The contours in both CDM simulations extend $\sim 75$ kpc behind the LMC, and are $\sim$ 90 kpc wide. The FDM wake retains its characteristic stochasticity, though in the Infall case, the FDM response in velocity dispersion is significantly weaker than even the no-self-gravity CDM simulation, as the FDM contour is $\sim 30$ kpc thinner and shorter than in CDM. 

The profile plots in Figure \ref{fig:start_dm_radDisp_profile} illustrate the dispersion profile across the wakes (in the $x$-direction). The peak dispersion is \textit{slightly} ($\delta \sigma_z \sim 0.01$) higher in the self-gravity-on case. The peak of the FDM wake is now much weaker than either CDM simulation, reaching $\delta \sigma_z \sim 0.23$ as opposed to $\sim 0.29$ in CDM. 

Taken together, Figures \ref{fig:int_dm_radDisp_profile} and \ref{fig:start_dm_radDisp_profile} show that FDM wakes are dynamically colder overall than CDM wakes. This can be explained by considering how FDM granules react to a gravitational potential. FDM particles collect into the characteristic granules that have a size of approximately the de Broglie wavelength. When the gravitational potential changes significantly on a scale comparable to or smaller than a granule, gravity becomes less effective at doing work on the granule \citep{khlopov_gravitational_1985}. This reduces the effectiveness of the LMC at heating the wake, and produces an FDM wake with lower dispersion than a CDM wake.

The $\sim 20\%$ reduction in the velocity dispersion response of the FDM wake compared to CDM is consistent across both the Infall and Fiducial orbit cases. This result suggests that DF wakes in FDM will be $\sim 20\%$ colder than in CDM independent of the density of the medium or speed of the perturber. 

\subsection{Velocity Divergence} \label{subsec:dm_div}

\begin{figure*}
    \centering
    \includegraphics[width=1.0\textwidth]{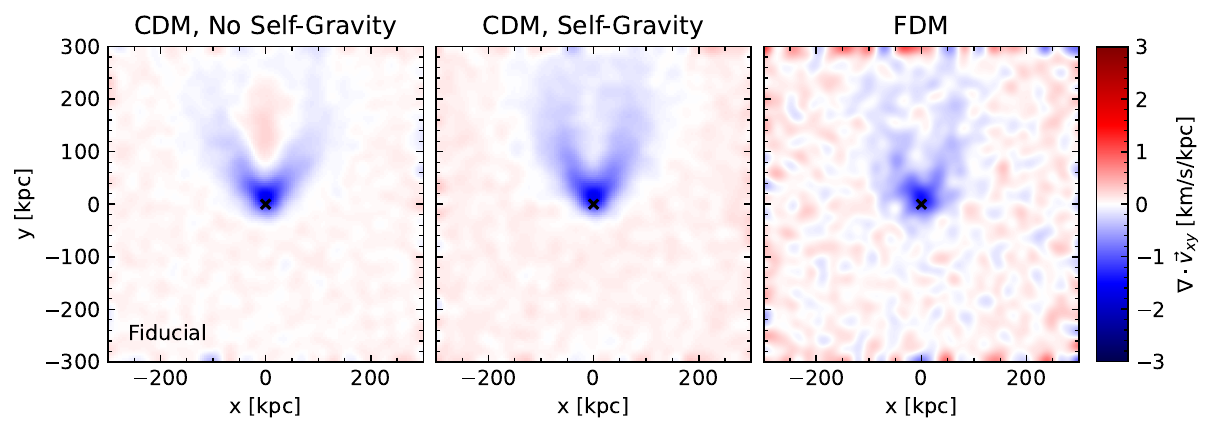}
    \caption{Divergence of the projected $x$-$y$ velocity field for the Fiducial case. See Section \ref{subsec:DM_disp} for how this is calculated. Regions of converging flows (blue, negative $\nabla \cdot \vec{v}$) clearly identify the wake boundaries and velocity structure in each simulation. The self-gravity-off CDM simulation in the left panel shows only the effect of the LMC on the particle velocities: Particles that pass close to the center of the LMC potential are deflected most strongly, producing a region of highly converging flow directly behind the LMC. After they are deflected, particles continue on relatively straight paths in the absence of wake self-gravity, producing a region of diverging flow at the center of the wake. Meanwhile, strongly deflected particles continue to move towards the box edges, passing near particles that have been deflected less strongly, creating a converging flow at the wake boundary. Turning on wake self-gravity (center panel) means that strongly-deflected particles continue to feel a force towards the wake after they pass the LMC. This eliminates the diverging region behind the LMC that is seen in the no-self-gravity case, and keeps the wake boundaries narrower. The net result is the effect seen in Figures \ref{fig:int_dm_density} and \ref{fig:int_dm_radDisp}: a denser, more coherent wake. The FDM simulation on the right shows a similar overall structure to the self-gravity-on CDM case, however, the granularity of the FDM results in a less smooth signature in the velocity divergence: The wake boundaries are less smooth, and some regions of converging flow can be seen inside the wake. Together with Figure \ref{fig:int_dm_radDisp}, it is clear that the FDM velocity field is not as coherently affected by the LMC as the CDM velocity field.}
    \label{fig:int_dm_velDiv}
\end{figure*}

\begin{figure*}
    \centering
    \includegraphics[width=1.0\textwidth]{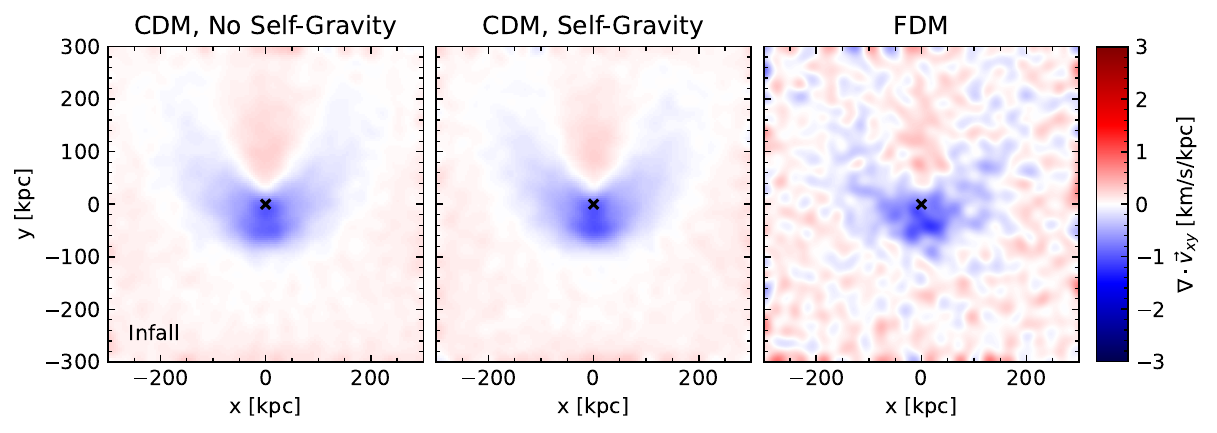}
    \caption{Same as Figure \ref{fig:int_dm_velDiv}, but for the Infall case after 2 Gyr of evolution. As in Figure \ref{fig:int_dm_velDiv}, the divergence traces the boundary of the wake. However, in all cases, the converging regions behind the LMC illustrate a wider opening angle than in the Fiducial case. This results directly from the lower wind speed for the Infall case. In the Infall case, the wake's self-gravity is not sufficient to erase the diverging flow at the center of the wake, although it is slightly narrower with self-gravity on, i.e. at $y=150$ kpc, the diverging region is $\sim 160$ kpc wide with self-gravity, compared to $\sim 180$ kpc wide without self-gravity. Meanwhile, the differences between the FDM simulation and its CDM counterparts are similar to those in Figure \ref{fig:int_dm_velDiv}: the resistance of FDM granules to strong accelerations washes out the coherent velocity signature seen in CDM.}
    \label{fig:start_dm_velDiv}
\end{figure*}

To help explain our results for the wake density and velocity dispersion, we also plot the divergence of the bulk velocity field to study how the particles are deflected by the LMC and the self-gravity of the wake. We again begin with the same 120-kpc wide slice about the $z$-midplane, and bin the particles/cells into an $x$-$y$ grid, this time with 4 kpc bins. In each bin, we calculate the mean $x$ and $y$ velocity components, leaving us with a 150x150 grid of 2-D velocity vectors. We then calculate the divergence of this 2D velocity field. Finally, we apply a Gaussian kernel of $\sigma = 12$ kpc to the result to reduce noise. 

Figure \ref{fig:int_dm_velDiv} shows the resulting
divergence maps for the Fiducial simulations. The wake signature shows up as regions of negative divergence (blue) tracing where the bulk flow of wind particles is converging. In all DM models, the region of strongest convergence is directly behind the LMC, where its gravity most strongly deflects particles. After being deflected, the particles cross the undeflected wind at larger impact parameters and create a region of converging flow that effectively traces the boundary of the wake. The crossing streams of particles behind the LMC produce the enhancement in the velocity dispersion seen in Figure \ref{fig:int_dm_radDisp}.

Comparing the CDM simulations, we can now pinpoint the effect that the wake's self-gravity has on the particle kinematics and wake structure. In the simulation without self-gravity, the particles deflected by the LMC simply continue on straight paths, creating a region of diverging velocities immediately downstream of the LMC. When self-gravity is turned on, the pull of the wake continues to deflect particles towards the center of the box, eliminating the diverging region, narrowing the wake boundaries, and enhancing the wake's density and velocity dispersion. 

As in Figure \ref{fig:int_dm_radDisp}, the FDM reacts less coherently to the LMC in velocity space, and the granularity persists in this kinematic signature. Despite the FDM simulations having self-gravity, the FDM wake shows regions of diverging velocity within the wake of a similar size scale to the granules in velocity space.

Figure \ref{fig:start_dm_velDiv} illustrates the velocity divergence of the wake produced in the Infall case for all three DM models. At this lower wind speed, the particles can be deflected significantly \textit{before} they reach the LMC center. Overall, particles are deflected more strongly when the wind speed is reduced, leading to wider wake boundaries where these strongly deflected particles cross over the undisturbed wind. The larger deflection angles make it more difficult for the wake's gravity to keep the deflected streams together, and while the self-gravity results in a slight narrowing (by $\sim 20$ kpc at $y=150$ kpc) of the downstream diverging region, it is not sufficient to eliminate the diverging flow behind the LMC. Larger deflections also cause a stronger velocity dispersion signature in the Infall case (see Figure \ref{fig:start_dm_radDisp}) compared to the Fiducial case (see Figure \ref{fig:int_dm_radDisp}).

In the Infall case, the wake boundaries are much less clear in FDM, just as they are in the Fiducal case (see Figure \ref{fig:int_dm_velDiv}). While the converging region in front of the LMC is clear, the granularity almost entirely washes out the wake boundaries. 

Overall, the velocity divergence illuminates several results from the previous two sections. In the Fiducial case, the wake's self-gravity eliminates the diverging flow in the center of the wake, raising the wake's density by $10\%$ and increasing the distance the wake takes to decay by $\sim 35\%$. In the Infall case, the diverging region remains regardless of the wake's self-gravity due to the increased deflection angles of the particles, which explains why the CDM Infall wakes look very similar regardless of self-gravity. FDM's granularity in the velocity divergence persists across both wind speeds and densities, showing that FDM does not react as coherently to a perturber as CDM. In turn, FDM wakes have lower velocity dispersions than their CDM counterparts. 

\section{Stellar Wakes} \label{sec:star_wakes}

\begin{figure*}[ht!]
    \centering
    \includegraphics[width=1.0\textwidth]{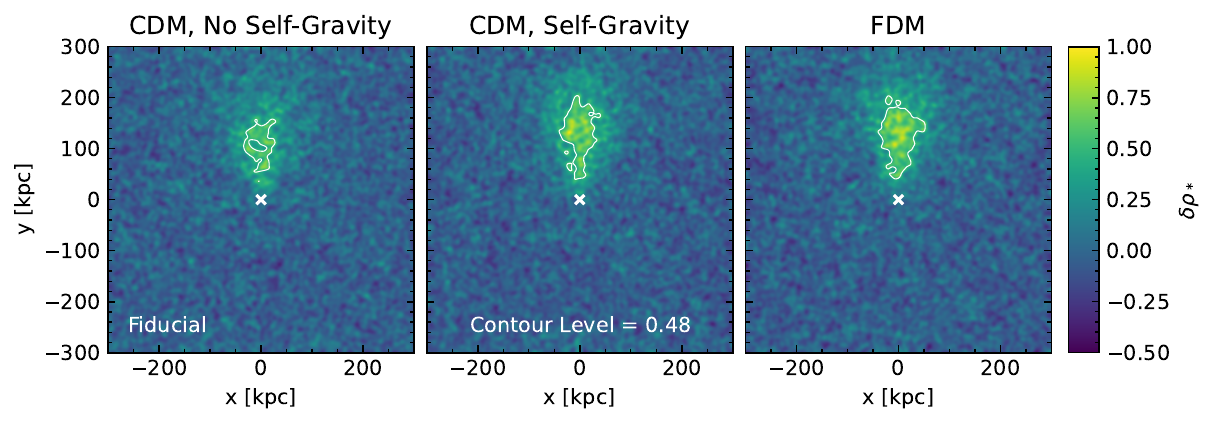}
    \caption{Projected overdensity (similar to Figures \ref{fig:int_dm_density} and \ref{fig:start_dm_density}) of stars in the Fiducial simulations after $t=0.7$ Gyr of integration. As in Figures \ref{fig:int_dm_density} and \ref{fig:start_dm_density}, a white contour in each panel traces the region with a density greater than the half-max of the CDM simulation with self-gravity (center). The CDM simulation without self-gravity (left) shows the stellar wake that is present due to \textit{only} the passage of the LMC, i.e. we would expect this wake to form in the stellar halo regardless of the behavior (or existence) of the DM wake. When compared to the stellar wake with CDM self-gravity (center), we see that the entire wake is not enclosed by the contour in the absence of self-gravity (left), i.e. there is a region in the center of the wake without CDM self-gravity that drops below $\delta \rho_* = 0.48$. This demonstrates that adding the gravity of the DM wake enhances the density of the stellar wake, which we quantify in Figure \ref{fig:int_stars_density_tavg}. Additionally, the contoured region extends $\sim 50$ kpc farther behind the LMC when the DM wake's gravity is included. Comparing the CDM simulation with self-gravity to the FDM simulation (right), the stellar wake's density looks similar in both of these DM models.}
    \label{fig:int_stars_density}
\end{figure*}

\begin{figure}
    \centering
      \includegraphics[width=1.0\columnwidth]{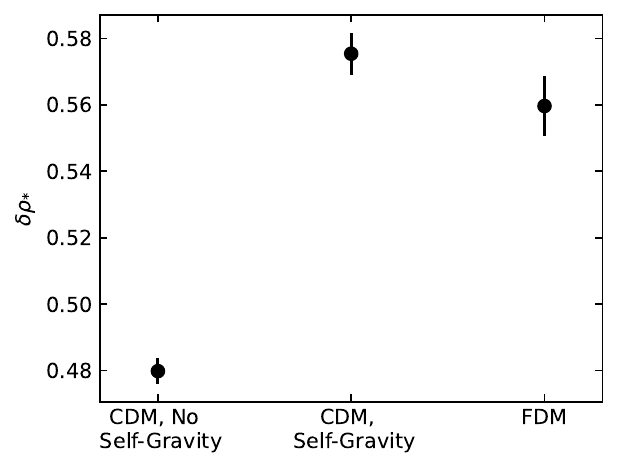}
    \caption{Median densities of the stellar wakes in the Fiducial simulations, averaged over 100 Myr of evolution from $t=0.6-0.7$ Gyr. Error bars show the standard deviation of the median density during this time period. The stellar wake's density is increased by $\delta \rho_* \sim 0.1$ when the gravity of the DM wake is added. The stellar wake is less dense in response to an FDM wake than a CDM wake, but this difference is small, only $\delta \rho_* \sim 0.02$. }
    \label{fig:int_stars_density_tavg}
\end{figure}

\begin{figure*}
    \centering
    \includegraphics[width=1.0\textwidth]{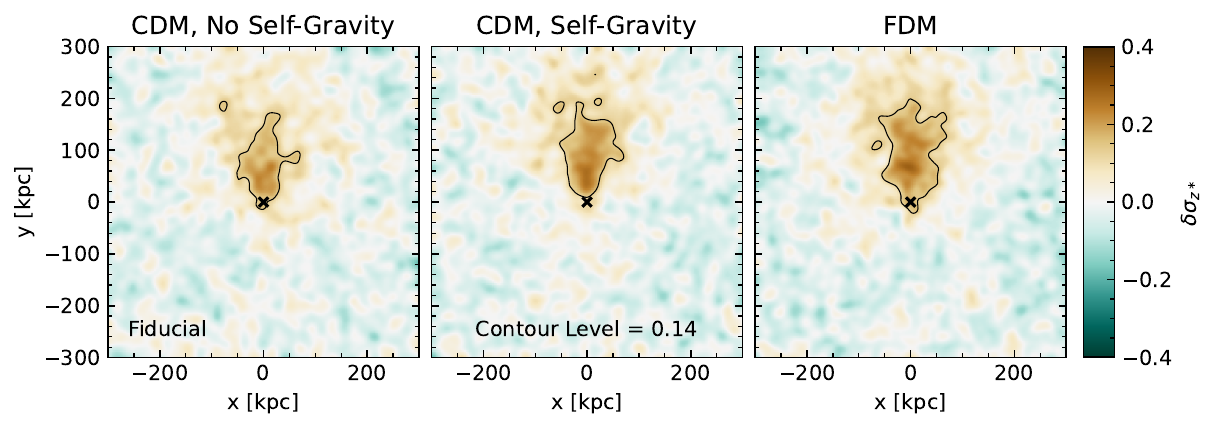}
    \caption{$z$-velocity dispersion of the stellar wakes in the Fiducial simulations (similar to Figure \ref{fig:int_dm_radDisp} but for the stars). Comparing the CDM simulations, we see that the addition of the DM wake's gravity widens the contour by $\sim 20$ kpc. While the contour extends roughly 180 kpc behind the LMC in both CDM simulations, the wake tapers more quickly without the DM self-gravity. Comparing the FDM simulation to the CDM simulation with self-gravity, the FDM wake tapers more slowly than either CDM simulation, though it reaches a similar maximum width to the CDM simulation with self-gravity.}
    \label{fig:int_stars_radDisp}
\end{figure*}
    
\begin{figure}
    \centering
    \includegraphics[width=1.0\columnwidth]{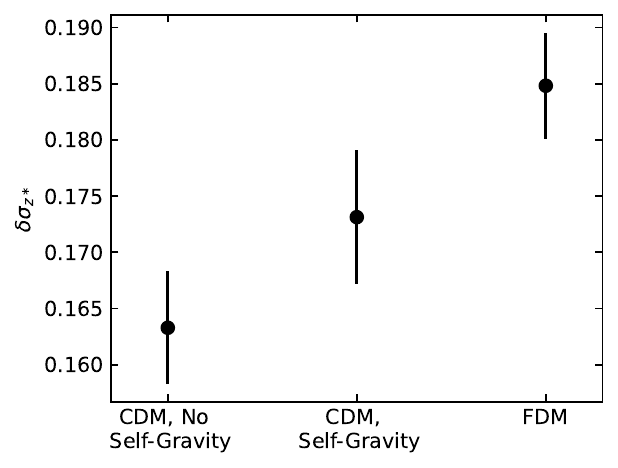}
    \caption{Time-averaged median enhancement of the $z$-velocity dispersion in the stellar wakes. The stellar wake's velocity dispersion is raised by $\delta \sigma_{z*} \sim 0.01$ when adding the gravity of the DM wake. The stellar wake is heated $\delta \sigma_{z*} \sim 0.01$ more by an FDM wake than a CDM wake.}
    \label{fig:int_stars_radDisp_tavg}
\end{figure}

Now, we turn our attention to the observable stellar counterpart of the LMC's wake. As a reminder, the stellar wind input parameters (see Table \ref{tab:star_wind_prop}) are meant to mimic the MW's stellar halo at 70 kpc, just as the DM initial conditions in the Fiducial orbit case match the MW's DM halo at 70 kpc (see Table \ref{tab:star_wind_prop}).

In $\S$ \ref{sec:DM_results}, we argued that the influence of DM self-gravity on the wake properties should be most observable at distances of 70-100 kpc. Extending this argument to the stellar wake, the best observational signatures of the stellar wake and the DM wake's influence on it should be between 70-100 kpc.

For this reason, we focus only on the Fiducial simulations in our discussion of stellar wakes. As with the DM wakes, we examine the density and velocity structure of the stellar wakes and identify signatures with which to confirm: 1) the presence of a stellar wake; 2) the presence of a DM wake; and 3) distinguishing features between a CDM or FDM wake. The observability of these signatures will be discussed further in Section \ref{subsec:obs}.

Figure \ref{fig:int_stars_density} shows the density structure of the stellar wakes in the Fiducial simulations. To make these plots, we use a procedure identical to Figure \ref{fig:int_dm_density}, with a single additional step of smoothing the resulting density fields with a Gaussian kernel with $\sigma = 4$ kpc. This additional smoothing is done to reduce the noise that results from sampling $\sim 100$ times fewer stars than DM particles. We again include a contour which encloses the region with overdensities higher than the half-max of the CDM simulation with self-gravity.

The left panel shows the stellar wake in the absence of DM self-gravity, i.e. the stellar wake that would form due to \textit{only} the passage of the LMC. The contour extends for roughly 150 kpc behind the LMC. In contrast, the center panel shows the stellar wake that forms when the stars feel the gravity from the CDM wake. 

The more striking difference is that the contour extends $\sim 50$ kpc farther behind the LMC than in the wake without self-gravity. This demonstrates that the DM wake's self-gravity holds the stellar wake together. Observationally confirming the existence of a stellar wake with $\delta\rho_*\gtrsim 0.6$ more than 150 kpc behind the LMC would provide strong evidence for the existence of a DM DF wake behind the LMC. Comparing the CDM simulation with self-gravity to the FDM simulation (right), however, reveals little difference in the 
density of the stellar wakes formed under the gravity of different DM particles. 

The profile-style plots (e.g. Figures \ref{fig:int_dm_density_profile} and \ref{fig:int_dm_radDisp_profile}) become very noisy when made with star particles due to the $\sim$ 100 times smaller sample sizes in each bin (with respect to DM). Instead, to compare the overall strength of the stellar response in each DM model, we compute an estimate of the overall wake density (Figure \ref{fig:int_stars_density_tavg}), time-averaged over five snapshots spanning 100 Myr of evolution. In detail, for each of the five snapshots, we compute a 2-D histogram of the quantity of interest (exactly as in Figure \ref{fig:int_stars_density} for the density, or in Figure \ref{fig:int_stars_radDisp} for the dispersion). For each histogram, we select bins with values that are over half that of the maximum bin, then take the median of these. The values reported in Figures \ref{fig:int_stars_density_tavg} are \ref{fig:int_stars_radDisp_tavg} the time-average and standard deviation of the medians.

\begin{figure*}
    \centering
    \includegraphics[width=1.0\textwidth]{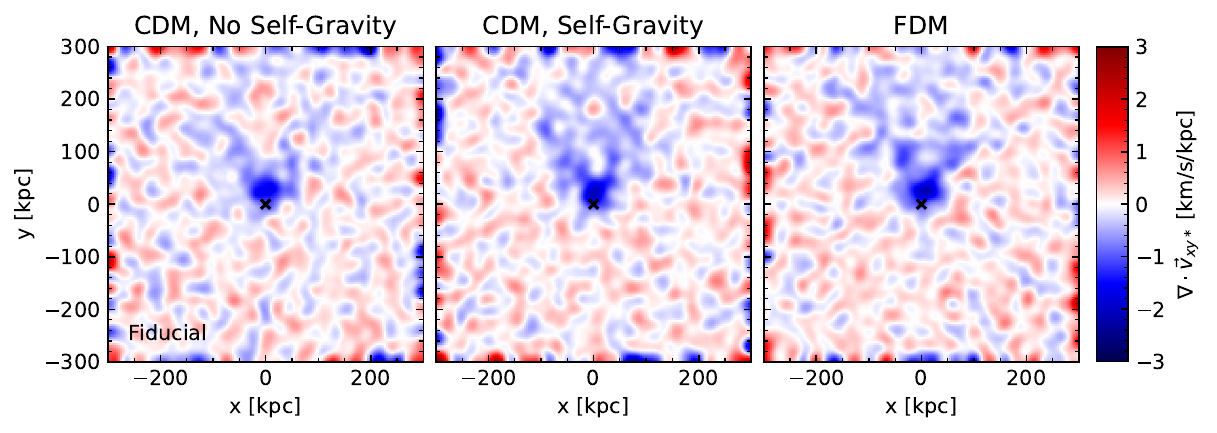}
    \caption{Divergence of the $x$-$y$ velocity field for the stars in the Fiducial simulations. As in Figure \ref{fig:int_dm_velDiv}, the star particles in the no-self-gravity CDM simulation (left) converge just behind the LMC and then continue on relatively straight paths, forming a divergent region farther downstream. Turning on DM gravity in the other panels largely eliminates this diverging region just as it does for the DM. This demonstrates how the DM wake enhances the density of the stellar wake and keeps it coherent for larger distances behind the LMC (see Figure \ref{fig:int_stars_density}, but there are no discernible differences between CDM and FDM here.}
    \label{fig:int_stars_velDiv}
\end{figure*}

Figure \ref{fig:int_stars_density_tavg} shows the time-averaged median density of the stellar wakes in each Fiducial simulation. The stellar wake reaches an overdensity of $\sim 0.48$ when the DM wake's gravity is not included, compared to $\sim 0.58$ when including the gravity of a CDM wake. The stellar wake in the FDM simulation reaches $\delta \rho_* \sim 0.56$, similar to the CDM with self-gravity case. 

In short, the gravity of a DM wake raises the density of the stellar wake by $\delta \rho_* \sim 0.1$, and extends the density response by $\sim 50$ kpc. CDM and FDM wakes do not leave significantly different signatures in the density of the stellar wake. 

Figure \ref{fig:int_stars_radDisp} shows the $z$-velocity dispersion of the stars in the Fiducial simulations, exactly as Figure \ref{fig:int_dm_radDisp} but for the star particles. The smoothing length is also increased to 9 kpc to mitigate the increased noise associated with the relatively low number of star particles. The velocity dispersion signature in the CDM simulation without self-gravity (left) is $\sim 20$ kpc narrower than when DM self-gravity is included (center and right). Additionally, when compared to the CDM simulation without self-gravity, the contour tapers more slowly in the CDM simulation with self-gravity and more slowly still in the FDM simulation.  

Figure \ref{fig:int_stars_radDisp_tavg} shows the time-averaged median enhancement in the $z$-velocity dispersion in the same fashion as Figure \ref{fig:int_stars_density_tavg}. A CDM wake's gravity raises the dispersion of the stellar wake by $\delta \sigma_{z*} \sim 0.01$ compared to when the DM wake's gravity is not present. Importantly, an FDM wake heats the stars more than a CDM wake: the velocity dispersion of the stars is $\delta \sigma_{z*} \sim 0.01$ higher in the FDM simulation than the CDM simulation with self-gravity. 

We plot the divergence of the $x$-$y$ velocity field to illuminate the density and kinematic structure of the stellar wakes (Figures \ref{fig:int_stars_density} and \ref{fig:int_stars_radDisp}) in Figure \ref{fig:int_stars_velDiv}. In the absence of the DM wake's gravity (left panel), we again see a region of converging flows immediately behind the LMC (blue), followed by a region of diverging flows (red) farther downstream as deflected stars pass by each other. Adding the gravity of the DM wake eliminates the diverging region just as it does for the DM particles, enhancing the density and velocity dispersion of the stellar wakes with DM self-gravity. The divergence of the stellar velocities looks similar between CDM with self-gravity and FDM. 

Altogether, the velocity dispersion enhancement of the stellar wake is slightly ($\sim 5\%$) higher in response to an FDM wake compared to a CDM wake. The only difference in the forces on the stars in both cases is caused by the differences in the density fields of the DM wakes. In Section \ref{subsec:DM_density}, we showed that FDM granules persist and are even strengthened inside of a DF wake. Therefore, we expect that the additional heating of the stellar wake in the FDM simulation is due to the scattering of stars by FDM granules. This so-called ``granule heating'' has been well-studied in other contexts (e.g. \citealt{hui_ultralight_2017}; \citealt{church_heating_2019}; \citealt{bar-or_relaxation_2019}; \citealt{bar-or_relaxation_2021}; \citealt{chavanis_landau_2021}; \citealt{dalal_dont_2021}; \citealt{dalal_not_2022}) and is a known property of FDM. In $\S$ \ref{subsec:FDM_mass}, we will discuss the role of granule heating and its dependence on FDM particle mass further.

Ultimately, we have demonstrated that the gravity of the DM wake plays an important role in shaping the response of the stars. Specifically, the gravity of the DM wake raises the overdensity of the stellar response by $\sim 10\%$ and extends the stellar wake's density response by $\sim 50$ kpc. The enhancement in the velocity dispersion within the stellar wake is $\sim 5 \%$ higher when CDM self-gravity is turned on, and $\sim 5 \%$ higher in FDM compared to CDM.

\section{Discussion} \label{sec:discussion}
In this section, we discuss the implications of our results in a wider context. We introduce a toy model for the observables of the stellar wake in $\S$ \ref{subsec:obs}, assess the sensitivity of the LMC's orbit to the choice of DM particle in $\S$ \ref{subsec:drag_forces}, and discuss the DM wake's mass and its impact as a perturber of the MW's dark halo in $\S$ \ref{subsec:wake_mass}. 

\subsection{Observational Predictions} \label{subsec:obs}

\begin{figure}
    \centering
    \includegraphics[width=1.0\columnwidth]{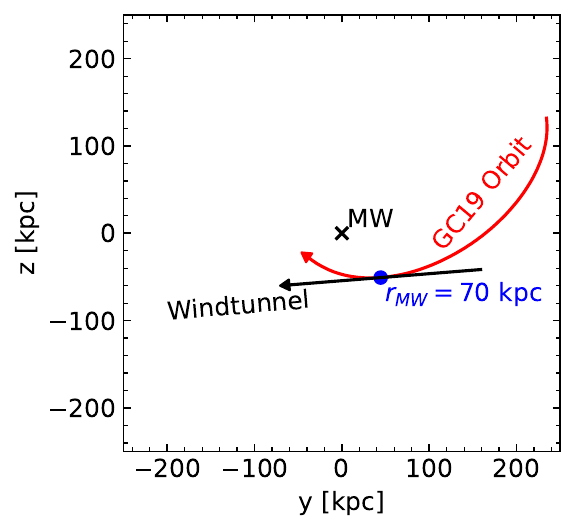}
    \caption{Schematic showing the location of the Fiducial windtunnel wake in our toy observational model in comparison to the LMC's orbit from the \citetalias{garavito-camargo_hunting_2019} reference simulation. As in Figure \ref{fig:orbit_cases}, we show both orbits in Galactocentric coordinates, projected onto the $yz$-plane. The MW center of mass is denoted by the cross, the \citetalias{garavito-camargo_hunting_2019} orbit is shown by the red line, and the black line shows the path of the LMC in our windtunnel simulations (equivalent to the $y$-axis in our simulation box). Arrows at the head of each path show the present-day location of the LMC. A blue dot shows the location at which the LMC is at a Galactocentric distance of 70 kpc, which is where we draw our Fiducial wind parameters. The path of the LMC in the windtunnel is tangent to the \citetalias{garavito-camargo_hunting_2019} orbit at this location.
    }
    \label{fig:obs_transform}
\end{figure}

In Section \ref{sec:star_wakes}, we presented three key predictions for the stellar wake. The gravity of a DM wake will: 1) enhance the overdensity of the stellar wake by roughly $10\%$; 2) extend the length of the stellar overdensity and kinematic response by a few tens of kpc; and 3) the velocity dispersion enhancement of the stellar wake will be mildly ($\sim 5\%$) higher in response to an FDM wake than a CDM wake. In this section, we assess the extent to which these results could be observable by introducing a toy model to approximate how our windtunnel wakes would be viewed from Earth. Using this toy model, we study the density and radial velocity dispersion of the stellar wake with the addition of simulated distance and radial velocity errors. 

To study how the stellar wake will appear when observed from Earth, we transform our ``windtunnel'' or simulation box coordinate system to Galactic ($l$,$b$,$r$) coordinates, in which the origin is the solar system barycenter, the $x$-axis points towards the Galactic Center, and the $z$-axis is normal to the Galactic plane. The LMC's path in the windtunnel is straight as opposed to a curved orbit, so we cannot exactly reproduce the appearance of the \citetalias{garavito-camargo_hunting_2019} wake on the sky, nor can we reproduce the effect of the collective response. 

However, we can carefully choose the transformation to ensure we are best-reproducing the orientation and location of the \citetalias{garavito-camargo_hunting_2019} wake in the region of sky where we want to make our observations. In this case, following \citetalias{garavito-camargo_hunting_2019} and our argument in $\S$ \ref{subsec:DM_density}, we want to focus our observations where the wake is at a Galactocentric distance of $70-100$ kpc. Therefore, our goal is to transform from windtunnel coordinates such that the straight windtunnel path is tangent to the LMC's orbit at a Galactocentric distance of 70 kpc, while the LMC itself is as close to its present-day location on the sky as possible. Our coordinate transformations are performed with \texttt{Astropy} version 4.2.1 (\citealt{astropy:2013}; \citealt{astropy:2018}; \citealt{astropy:2022}) and are described in Appendix \ref{apdx:coord_trans}.

\begin{figure*}
    \centering
    \includegraphics[width=1.0\textwidth]{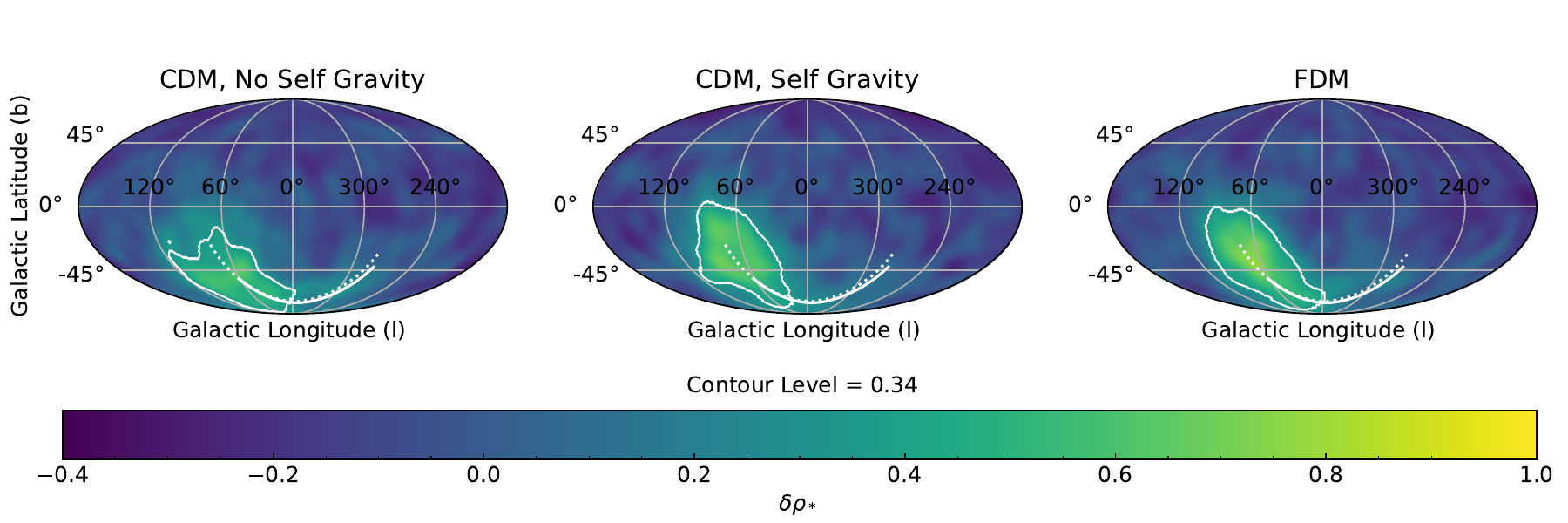}
    \caption{All-sky maps of the overdensity of stars in the Fiducial simulations after 0.7 Gyr of integration. Each panel shows a Mollweide projection in Galactic coordinates as defined by the \texttt{Astropy} convention, and includes stars with distances between 70 and 100 kpc. White lines show the paths of the LMC from our \citetalias{garavito-camargo_hunting_2019} reference simulation (dashed) and in our windtunnel (solid). As in Figure \ref{fig:int_stars_density}, the contour in each panel encloses the region with a response greater than the half-max of the CDM simulation with self-gravity, which is $\delta \rho_* > 0.34$ here. Panels show different DM models, with CDM without self-gravity on the left, CDM without self-gravity in the center, and FDM on the right. In all three panels, the wake appears as an overdensity covering nearly an entire octant of the sky, from $l \in \sim [0, 120]$ and $b \in \sim [-80, 0]$. The addition of self-gravity extends the contoured region to $b \approx 0$, while without self-gravity the wake decays to below the contour by $b \approx -20$. The extent of the stellar wake is similar in CDM and FDM when self-gravity is included.
    }
    \label{fig:allSky_density}
\end{figure*}

\begin{figure*}
    \centering
    \includegraphics[width=1.0\textwidth]{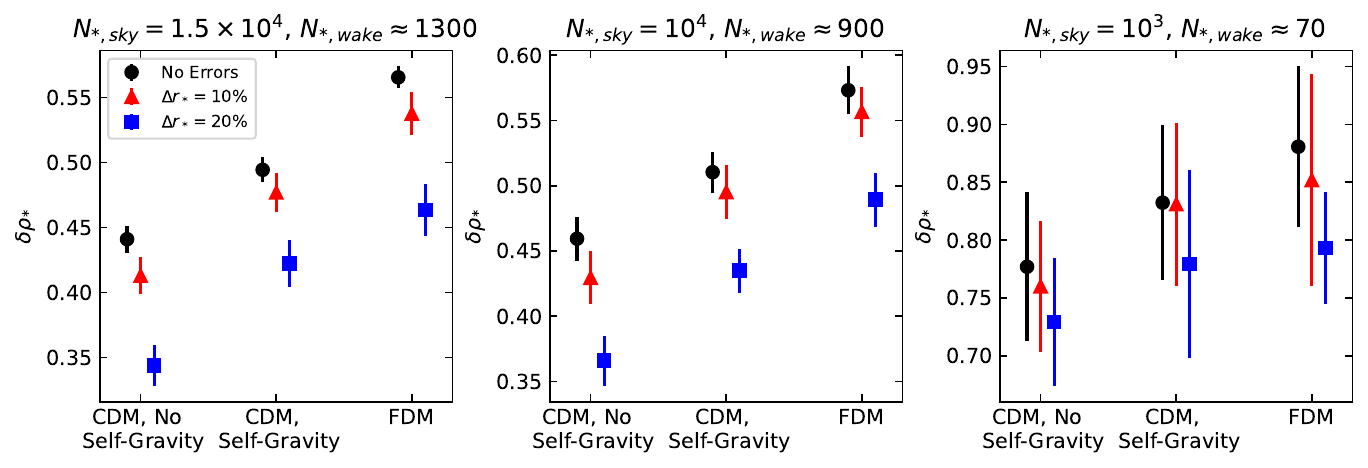}
    \caption{Time-averaged median wake densities (as in Figure \ref{fig:int_stars_density_tavg}) in the observable frame with observational errors. In each panel, we sample a different number of stars from the entire sky with distances between 70 and 100 kpc to compute the median wake density. Each point shows the average and standard deviation (computed via bootstrapping) of the median density across all sampled snapshots. Black circles show the simulation data with no errors. The red triangles (blue squares) have simulated distance errors of 10 (20) percent. With no observational errors at our highest sampling rate ($1.5 \times 10^4$ stars; left panel), the stellar wake is $\delta \rho_* \sim 0.05$ higher when the CDM wake's gravity is included, and the stellar wake in the FDM simulation is $\delta \rho_* \sim 0.07$ higher than in the CDM simulation with self-gravity. As observational errors increase, the measured overdensity of the wake decreases but the same trend between all three simulations remains. With $10^4$ stars (middle panel), the differences between DM models are lower but still distinguishable with 20\% distance errors. With only $10^3$ stars (right panel), the DM models are indistinguishable. 
    }
    \label{fig:avg_obsFrame_density_obsErrs}
\end{figure*}

The result of the coordinate transformation is shown in Figure \ref{fig:obs_transform}. We plot the LMC's orbit in the \citetalias{garavito-camargo_hunting_2019} reference simulation in Galactocentric coordinates in red as in Figure \ref{fig:orbit_cases}. In our toy observational model, the path of the LMC in the windtunnel is tangent to the LMC orbit from \citetalias{garavito-camargo_hunting_2019} at 70 kpc from the Galactic center, which is the distance that our Fiducial wind parameters are taken from.

To estimate how observational uncertainties affect our results, we also include Gaussian distance and radial velocity errors in our model. We choose two levels of errors, motivated by the performance of contemporary surveys. The distance errors are $10\%$ and $20\%$, typical for spectro-photometric distance measurements from DESI \citep{cooper_overview_2023} and the H3 survey \citep{conroy_mapping_2019}. For the radial velocity errors, we choose 1 and 10 km/s. Note that 1 km/s velocity errors reflect the performance of spectroscopic radial velocity measurements from DESI \citep{cooper_overview_2023}, \textit{Gaia} (\citealt{katz_gaia_2019}; \citealt{seabroke_gaia_2021}), and H3 \citep{conroy_mapping_2019}, while 10 km/s provides a reasonable worst-case scenario. 

With our toy model in-hand, we can now use it to study how the stellar wake might be observed. Figure \ref{fig:allSky_density} shows all-sky Mollweide projections (made with \texttt{Healpy}; \citealt{Gorski_2005_HEALPix, Zonca_2019_healpy})\footnote{https://healpix.sourceforge.io/} in Galactic coordinates of the overdensity of stars with distances of 70 - 100 kpc in the Fiducial simulations after 0.7 Gyr of evolution. The bin size is $1.16^o$ and the resulting density map is smoothed by a Gaussian kernel with $\sigma = 15^o$. Each panel corresponds to a DM model, with CDM without self-gravity on the left, CDM with self-gravity in the center, and FDM on the right. Each panel shows the path of the LMC in the windtunnel as the solid white line, and the LMC orbit from the reference simulation as the dashed white line; we see good agreement between the position of both paths on the sky. As in Figure \ref{fig:int_stars_density}, we enclose the region with an overdensity higher than 0.34 with a contour; this level is the half-maximum density of the CDM simulation with self-gravity.

In agreement with \citetalias{garavito-camargo_hunting_2019}, the stellar wake appears as an overdense region in the Galactic southeast, ranging from $l \sim$ 0 - 120, and $b \sim$ $-80$ - 0. Notably, the extension of the stellar wake owing to the DM wake's gravity is readily observable: while the stellar wakes in the CDM simulation with self-gravity and FDM simulation do not decay below $\delta \rho_*$ of 0.38 until $b \approx 0$, the stellar wake decays to this level by $b \approx -20$ in the simulation without self-gravity.

\begin{figure*}
    \centering
    \includegraphics[width=1.0\textwidth]{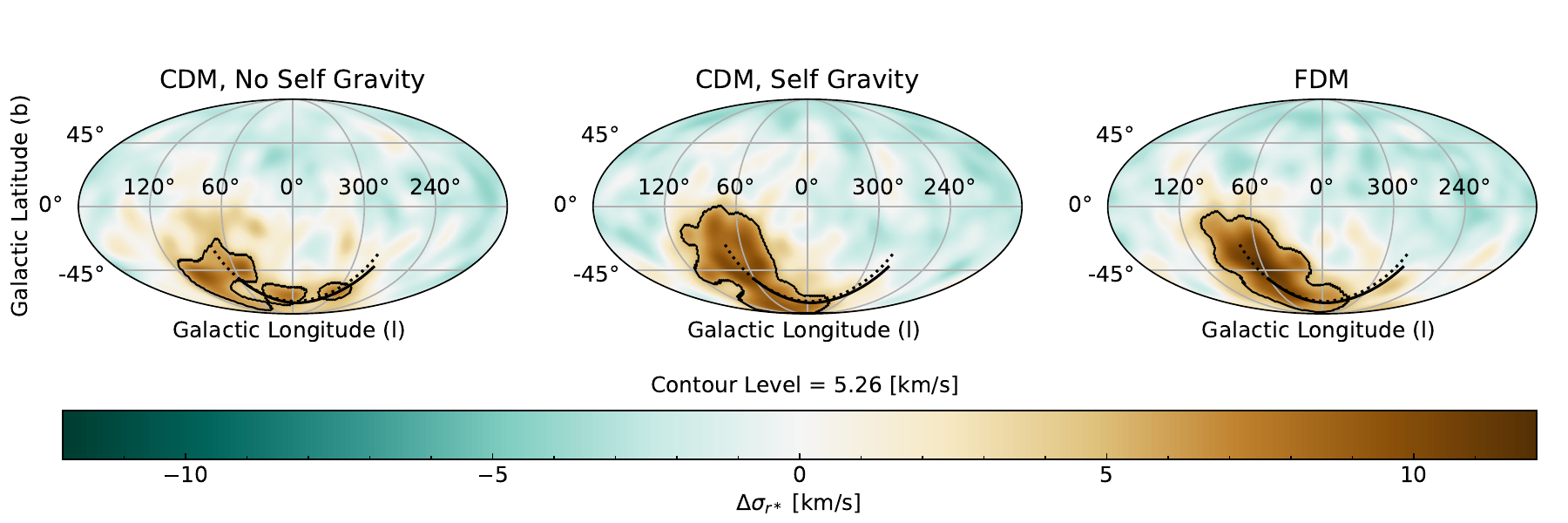}
    \caption{All-sky maps of the change in the radial velocity dispersion with respect to the shell-average, $\Delta \sigma_{r*} = \sigma_{r*} - \bar{\sigma_{r*}}$, for the Fiducial simulations after 0.7 Gyr. Similar to the simulation-box projection (Figure \ref{fig:int_stars_radDisp}), the wake signature is an enhancement of the stars' radial velocity dispersion. The location of the velocity dispersion response is similar to the location of the density response in Figure \ref{fig:allSky_density}. Without self-gravity, the wake decays below the contour by $b \approx -30$. Both FDM and CDM with self-gravity extend the stellar wake to $b \approx 0$.
    }
    \label{fig:allSky_radVel}
\end{figure*}

To quantify the differences in the strength of the response in this observed frame, we use the same procedure as in $\S$ \ref{sec:star_wakes}: we calculate the median wake density or velocity dispersion in bins that are higher than half of the maximum bin. For each simulation, we repeat this for five snapshots spanning 100 Myr, and then report the average and standard deviation of the medians from the five snapshots. In this section, we calculate the quantity of interest in on-sky bins as in Figures \ref{fig:allSky_density} and \ref{fig:allSky_radVel}. 

To estimate the number of stars that need to be observed to distinguish between the simulations, we also downsample the number of star particles, i.e. after adding simulated errors, we sample a fixed number of stars with distances between 70 and 100 kpc from the entire sky. Without downsampling, there are approximately $10^5$ stars with distances in this range based on the stellar wind density and the volume of the shell. For our plots, we choose three different levels of downsampling, selecting $1.5 \times 10^4$, $10^4$, and $10^3$ stars. These sampling levels correspond to selecting approximately 1300, 900, and 70 stars within the wake (i.e. inside the contours in Figure \ref{fig:allSky_density}), respectively. 

Figure \ref{fig:avg_obsFrame_density_obsErrs} shows the time-averaged median overdensity of the stellar wake between 70 and 100 kpc with different observational errors and sampling rates. The black circles show the mean and standard deviation of the median overdensity without any observational errors. The errorbars on each point are computed via bootstrapping, i.e. for each of the five snapshots we randomly sample errors and star particles 50 separate times such that the final reported median overdensity is over 250 samples.

With no errors and $1.5\times10^4$ stars, when we compare the CDM simulations with and without self-gravity, the stellar wake's overdensity increases by $\sim 0.05$ with self-gravity. In the observational frame, we now also see a further increase in density in the FDM simulation, with the FDM simulation reaching $\delta \rho_* \sim 0.07$ higher than the CDM simulation with self-gravity. Note that this is opposite to the trend we saw in $\S$ \ref{sec:star_wakes}, where the stellar wake was slightly less dense in FDM compared to CDM. In the observational model, we are now looking at a 30 kpc thick slice of the wake, as opposed to 120 kpc in $\S$ \ref{sec:star_wakes}, so this is most likely an effect of the viewing angle and distance selection of stars. When adding observational errors and reducing the number of stars to $10^4$, the differences between the simulations remain visible. Sampling only $10^3$ stars, however, is not sufficient to see the differences between the simulations. 

\begin{figure*}
    \centering
    \includegraphics[width=1.0\textwidth]{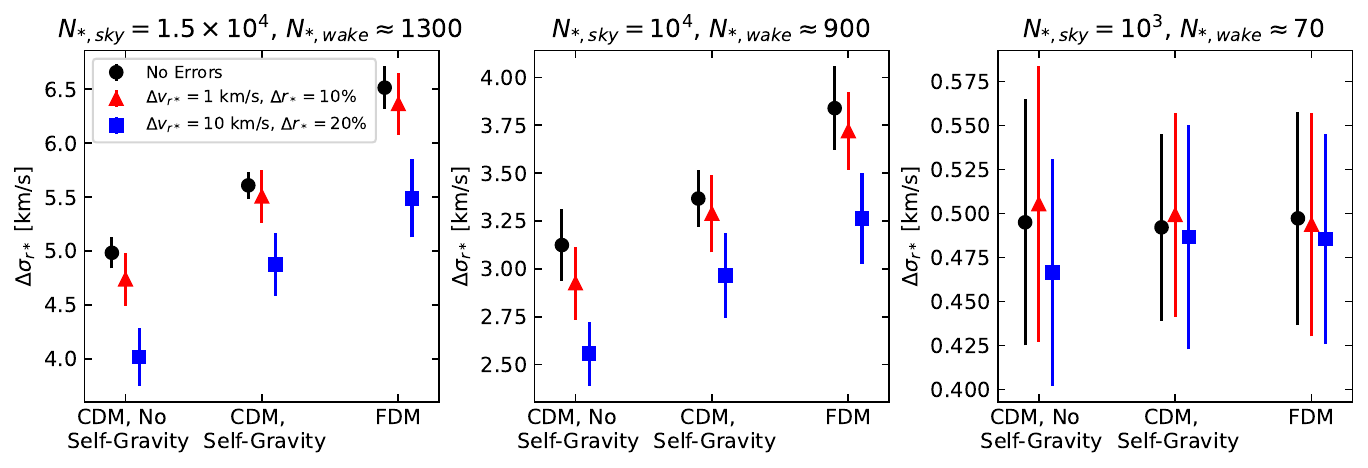}
    \caption{Time-averaged median radial velocity dispersion enhancement (same as Figure \ref{fig:avg_obsFrame_density_obsErrs} but showing $\Delta \sigma_{r*}$) of the stars in our toy observational model. In addition to distance errors, the red triangles (blue squares) have simulated radial velocity errors of 1 (10) km/s. With $1.5\times10^4$ stars and no observational errors, the velocity dispersion of the stellar wake is $\sim 0.6$ km/s higher than the simulation without self-gravity with the CDM wake's gravity, and $\sim 1.5$ km/s higher with the FDM wake's gravity. Adding observational errors reduces the measured velocity dispersion enhancement, though the differences between the simulations remain above 1-$\sigma$. With $10^4$ stars, the simulations are still distinguishable, but the differences reach just above 1-$\sigma$ with 10 km/s and 20\% distance errors. Sampling $10^3$ stars is insufficient to distinguish the simulations at any error level.} 
    \label{fig:avg_obsFrame_radDisp_obsErrs}
\end{figure*}

Figure \ref{fig:allSky_radVel} shows all-sky maps of the enhancement in the radial velocity dispersion of the stars in the same fashion as Figure \ref{fig:allSky_density}. In this plot, we report the velocity dispersion as the difference from the shell average $\Delta \sigma_{r*} = \sigma_{r*} - \bar{\sigma_{r*}}$. In all panels, the velocity response traces the location of the density response well. The addition of the DM wake's gravity extends the length of the velocity response, as it decays to below 5.26 km/s by $b\approx -30$ in the CDM simulation without self-gravity, compared to $b\approx0$ in both simulations with the DM self-gravity. 

It is also worth mentioning that we expect an increase in both the longitudinal and latitudinal velocity dispersion in the wake. At these distances (70-100 kpc), we measure this increase to be approximately 0.03 mas/yr. For our purposes of distinguishing between DM models, the qualitative differences between the simulations are the same as for the radial velocity dispersion so we do not elaborate on the proper motions here for brevity.

Figure \ref{fig:avg_obsFrame_radDisp_obsErrs} shows the the median velocity response averaged over 100 Myr with observational errors and different numbers of stars, similar to Figure \ref{fig:avg_obsFrame_density_obsErrs}. Here, we see the same trend in the observational frame that we did in the simulation box frame in $\S$ \ref{sec:star_wakes}: with $1.5 \times 10^4$ stars, the velocity dispersion enhancement in the stellar wake is lowest ($\sim 5.0$ km/s) without a DM wake's gravity, higher in response to a CDM wake ($\sim 5.6$ km/s), and highest in response to an FDM wake ($\sim 6.5$ km/s). The addition of observational errors does not affect this trend, i.e. the simulations are still distinguishable with the largest errors we consider. $10^4$ stars is also enough to distinguish the simulations, though the differences between CDM with self-gravity and FDM become close to 1-$\sigma$ with 10 km/s radial velocity errors and 20\% distance errors. The differences between the simulations are not visible while sampling only $10^3$ stars. 

With the caveat that our observational framework is only a toy model, we find that the general results reported in $\S$ \ref{sec:star_wakes} still hold. In particular, we have demonstrated several important qualitative results: Distinguishing the strength of the density and kinematic response of the stellar wake between DM models should be possible with $\geq10^4$ stars across the entire sky ($\gtrsim 900$ stars within the wake) with distances between 70 and 100 kpc. This sampling rate corresponds to a number density of $3.6\times10^{-3}$ kpc$^{-3}$ which agrees with the number density of stars that \citetalias{garavito-camargo_hunting_2019} reported is required to confidently detect the wake. In other words, if we observe enough stars to detect the wake, we have enough stars to distinguish between the DM models considered here. 

Provided this sampling rate is achieved, we find that the telltale sign of the presence of a DM wake is the length of the response, as both the density and velocity dispersion responses are lengthened by over $20^o$ on the sky when the self-gravity of the DM wake is included. Additionally, we find that differences in the kinematics of the stellar wake between a CDM and FDM universe are still visible when accounting for the viewing perspective and observational errors. As also reported by \citetalias{garavito-camargo_hunting_2019}, we find that the increased velocity dispersion is a characteristic signature of the wake that differentiates it from cold substructure such as stellar streams. Ultimately, these results demonstrate that kinematic information is crucial when making observations of DF wakes, both for detecting the wake and inferring the nature of its DM component. 

\subsection{Dynamical Friction Drag Forces and the LMC's Orbit} \label{subsec:drag_forces}

\begin{figure*}
    \centering
    \includegraphics[width=0.495\textwidth]{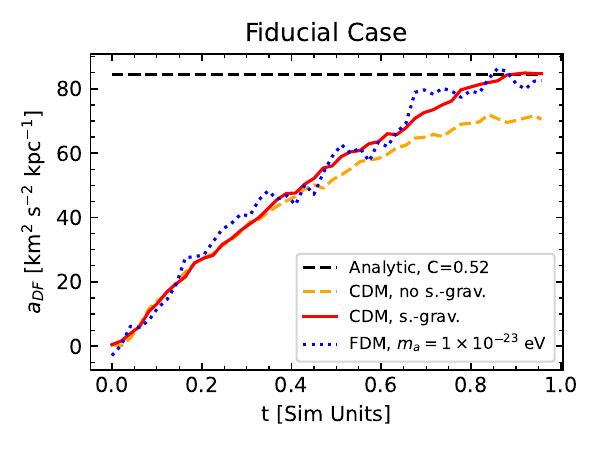}
    \includegraphics[width=0.495\textwidth]{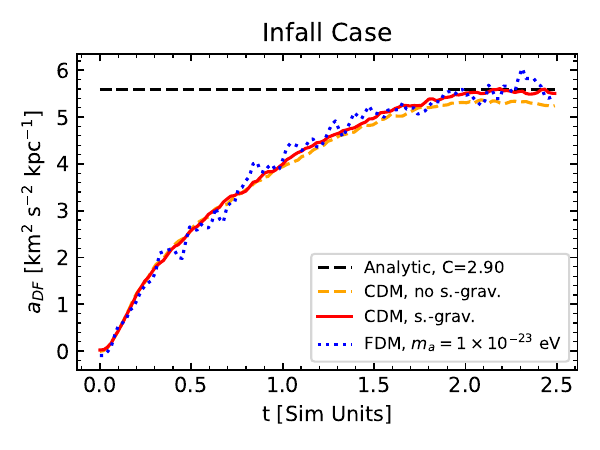}
    \caption{DF accelerations vs. time. The CDM simulations with and without self-gravity are shown by the solid red and dashed orange lines, respectively. The primary FDM mass ($m_a=10^{-23}$ eV) is the blue dotted line. The black dashed line in each panel shows the analytic acceleration calculated from the classical Chandrasekhar formula as described in the text. In both orbit cases, the measured acceleration increases in strength with time as the wake forms. \textit{Left panel}: The Fiducial orbit case. CDM produces a $\sim 10\%$ stronger acceleration when self-gravity is turned on, consistent with the enhancement in the wake's density and longevity when self-gravity is included. FDM produces a similar acceleration to the self-gravity-on CDM wake. The steady-state drag force is similar to the analytic approximation when ln$(\Lambda)$ is calculated using $C=0.52$. \textit{Right panel}: The Infall orbit case. The reduced effect of DM self-gravity in the Infall scenario is apparent in the similar accelerations between the two CDM simulations. Again, the FDM traces the CDM acceleration well. The steady-state drag force matches the analytic approximation when ln$(\Lambda)$ is calculated using $C=2.90$. The drag force in the Fiducial case is much higher than the drag force in the Infall case, as the background density is roughly 2 orders of magnitude larger, while the wind velocity is only a factor of $\sim3$ larger. Overall, the FDM drag force matches the CDM drag force well, consistent with the predictions of \citet{lancaster_dynamical_2020} with respect to DF on the LMC from an FDM wake.}
    \label{fig:drag}
\end{figure*}

In this section, we compare the behavior of the DF drag force felt by the LMC due to the DM wakes in our simulations and discuss the impact of DM microphysics on the LMC's orbit. 

To determine the acceleration due to DF in our simulations, we calculate the $y$-component of the gravitational acceleration that would be felt by a constant-density sphere 5 kpc in radius at the center of the box due to all DM particles in the simulation. When done at each timestep, this gives us an approximation of the DF acceleration felt by the LMC as a function of time.

Additionally, we calculate the expected DF acceleration using the classic formula from \citet{chandrasekhar_dynamical_1943}:

\begin{equation}
    a_{DF} = \frac{4\pi^2 M G^2 \bar{\rho} \rm{ln}(\Lambda)}{\sigma^2}\frac{1}{2X^2}\left[ \rm{erf}(X) -\frac{2X}{\sqrt{\pi}} e^{-X^2}\right] ,
\end{equation} \label{eqn:chandra_df}

where erf is the error function and
\begin{equation}
    X=v/\sqrt{2}\sigma.
\end{equation}

In these equations, we use the input wind parameters and LMC mass, i.e. $M$ from Table \ref{tab:gal_models}, and $\bar{\rho}$, $v$, and $\bar{\sigma}$ from Table \ref{tab:dm_wind}.

For the Coulomb logarithm, we follow \citet{van_der_marel_m31_2012}, \citet{patel_orbits_2017}, and \citetalias{garavito-camargo_hunting_2019}, using

\begin{equation}
    \text{ln}(\Lambda) = \text{max}\left[L, \text{ln}(r/Ca)^\alpha \right]
\end{equation} \label{eqn:coulog}

where $r$ is the distance between the satellite and its host, $a$ is the satellite's scale radius, and $L=0$ and $\alpha=1$, and $C$ are constants. Here, $r$ is the separation between the LMC and MW at the point in the reference simulation that we base our wind parameters on (70 kpc for the Fiducial case and 223 kpc for the Infall case), and $a$ is the LMC's scale radius from Table \ref{tab:gal_models}. For $C$, we pick values such that the analytic DF acceleration roughly agrees with the measured acceleration when the wake reaches the end of the box. For the Fiducial wind, this is $C=0.52$, and for the Infall wind, $C=2.90$. 

\begin{figure*}
    \centering
    \includegraphics[width=1.0\textwidth]{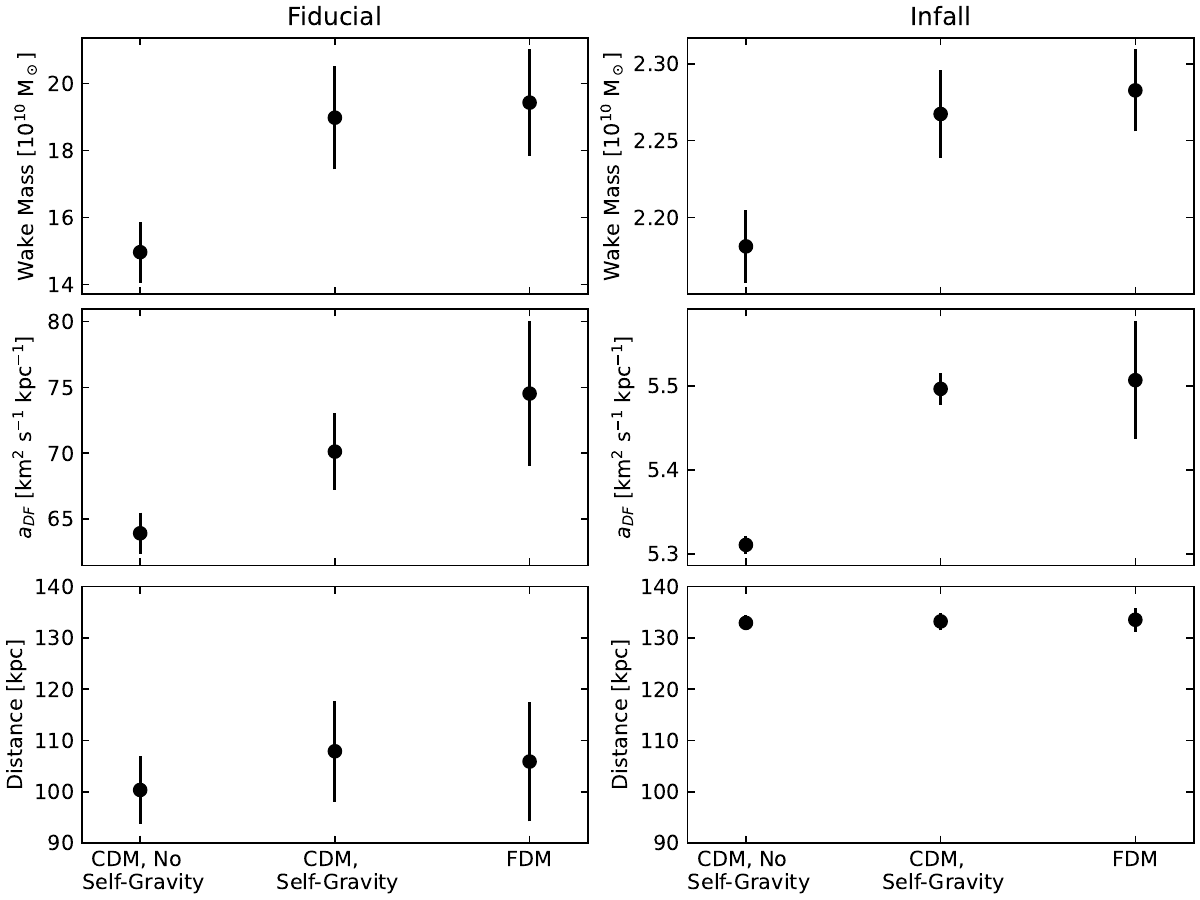}
    \caption{Wake masses in our simulations, calculated as described in the text. In all panels, quantities are averaged over five snapshots spanning 100 Myr, and the points show the mean and standard deviation of the quantity of interest. The left column of panels shows the Fiducial wind simulations from 0.6-0.7 Gyr of evolution, and the right column shows the Infall wind simulations from 1.9-2 Gyr of evolution. The top row shows the wake masses. In the Fiducial case, the wake masses are roughly that of the LMC's virial mass at infall, while in the Infall case, they are roughly an order of magnitude less. In both wind cases, the CDM wake without self-gravity is of order $10\%$ less massive than the CDM wake with self-gravity. Both wakes with self gravity have similar masses. The middle row of panels shows the DF acceleration on the LMC due to the wake during at the same times as we measure the wake mass. See $\S$ \ref{subsec:drag_forces} and Figure \ref{fig:drag} for a further discussion of the drag forces. The bottom row of panels shows the distance at which a point object of the mass in the upper row would produce the gravitational acceleration in the middle row, for each simulation. Thus, the Fiducial wake acts like an LMC-mass perturber trailing the LMC at a distance of $\sim100$ kpc, while the Infall wake acts as a perturber with $\sim 1/10$ the mass of the LMC at a distance of $\sim 135$ kpc.
    }
    \label{fig:wake_masses}
\end{figure*}

Figure \ref{fig:drag} shows the measured and analytic DF accelerations for our simulations, with the Fiducial wind case in the left panel and the Infall wind case on the right. In each simulation, the strength of the drag increases with time as the wake forms, before plateauing once the faster-moving wake particles begin to wrap though the box. Overall, the drag from the Fiducial wake is slightly more than an order of magnitude stronger than the drag from the Infall wake, which aligns with the $\rho/v^2$ scaling expected from Equation \ref{eqn:chandra_df}. 

In the Fiducial case, we see that the reduction in wake size and density when self-gravity is removed translates to a weaker drag force - the acceleration is $\sim 10\%$ weaker in the CDM simulation without self-gravity vs. with self-gravity. Meanwhile, the behavior of the FDM drag is consistent with the predictions of \citet{lancaster_dynamical_2020}, who calculated that the time-averaged drag force on the LMC should be well-approximated by classical DF (i.e. with non-interacting background particles). In the Infall case, we see closer agreement between the two CDM simulations, as the effect of the wake's self-gravity is diminished at this lower wind speed.

Ultimately, our result that both DM models produce a similar drag force regardless of the wind speed and density (when DM self-gravity is included) implies that the LMC's orbit would not be impacted by the assumption of a CDM vs. FDM universe. 

\begin{table*}[t]
    \centering
    \caption{Summary of our additional simulations for $\S$ \ref{sec:caveats}, listing the properties for each simulation. We list: The simulation category along with the subsection where the simulations are discussed; the dark matter model with the FDM particle mass if applicable; whether DM self gravity is enabled; the number of DM resolution elements (\textit{N}-body particles for CDM, grid cells for FDM); the orbit/DM wind case (see Table \ref{tab:dm_wind}); and how the simulation differs from those described in Table \ref{tab:dm_wind}.
    }\label{tab:adtl_sims}
    \begin{tabular}{c|c c c c c c }
        \hline
        \hline
        Category & DM Model & DM Self-Gravity & Particles or Cells & DM Wind & Notes \\
        \hline
        Low-Mass & FDM, $m_{a}=2.5 \times 10^{-24}$ eV & Yes & $256^3$ & Fiducial & \tablenotemark{a}\\
        FDM, $\S$ \ref{subsec:FDM_mass}\ & FDM, $m_{a}=2.5 \times 10^{-24}$ eV & Yes & $256^3$ & Infall & \tablenotemark{a} \\
        \hline
        Alternative LMC & CDM & No & $10^8$ & Fiducial & \tablenotemark{b} \\
        Masses, $\S$ \ref{subsec:LMC_mass} & CDM & No & $10^8$ & Fiducial & \tablenotemark{c} \\
        \hline
        Low Stellar & CDM & Yes & $10^8$ & Fiducial & \tablenotemark{d} \\
        Dispersion, $\S$ \ref{subsec:sig_star} & FDM, $m_{a}=10^{-23}$ eV & Yes & $1024^3$ & Fiducial & \tablenotemark{d} \\
        \hline
        \hline
    \end{tabular}
    \tablenotetext{a}{FDM particle mass is reduced by a factor of four.}
    \tablenotetext{b}{Uses the ``Light LMC,'' see Table \ref{tab:lmc_models}.}
    \tablenotetext{c}{Uses the ``Heavy LMC,'' see Table \ref{tab:lmc_models}.}
    \tablenotetext{d}{The stars in this simulation have a velocity dispersion of $\bar{\sigma}_*=30.0$ km/s.}
\end{table*}

\begin{table}[]
    \caption{Summary of our additional LMC mass models, also from \citetalias{garavito-camargo_hunting_2019}. We list: our galaxy model; the corresponding galaxy model in \citetalias{garavito-camargo_hunting_2019}; $M$, the total mass of the profile (if it were integrated to infinity); and $a$, the scale radius.} \label{tab:lmc_models}
    \centering
    \begin{tabular}{c c c c}
        \hline
        \hline
         Galaxy Model & \citetalias{garavito-camargo_hunting_2019} Model & $M$ [M$_\odot$] & $a$ [kpc] \\
         \hline
         Light LMC & LMC2 & $1.0 \times 10^{11}$ & 12.7 \\
         Heavy LMC & LMC4 & $2.5 \times 10^{11}$ & 25.2 \\
         \hline
         \hline
    \end{tabular}
\end{table}

\subsection{The Mass of the Wake}\label{subsec:wake_mass}

In this section, we calculate the mass of the DM wakes in our simulations, and develop a basic framework to understand the DM wake as a perturbation to the MW's DM halo.  

To calculate the wake mass in each of our simulations, we begin by defining a rectangular region that roughly contains the wake (i.e. that contains where $\delta \rho \geq 0.1$; $x \in [-100, 100]$, $y \in [-50, 300]$, $z \in [-100, 100]$ for the Fiducial wake; $x \in [-150, 150]$, $y \in [-50, 300]$, $z \in [-150, 150]$ for the Infall case). At a particular timestep, the wake mass is estimated by taking the difference between the total DM mass within the region at that timestep and the mass within the region at the start of the simulation, i.e the region's volume multiplied by $\bar{\rho}$ from Table \ref{tab:dm_wind}. As we have done throughout this work, when estimating the wake mass, we average over five snapshots spanning 100 Myr of evolution. 

The top row of Figure \ref{fig:wake_masses} shows the masses of all DM wakes in our simulations. The left panel shows the Fiducial wind after 0.7 Gyr of evolution, and the right panel shows the Infall wind after 2 Gyr of evolution. The mass of the Fiducial wake is roughly comparable to the LMC, while the mass of the Infall wake is roughly an order of magnitude lower. In both the Infall and Fiducial case, the FDM wake and CDM wake with self-gravity have similar masses, while the CDM wake without self-gravity is of order $10\%$ less massive than either wake with self-gravity.

To get a rough approximation of the impact of the DM wake as a perturbation to the MW's DM halo, we also calculate the distance at which an object with the wake's mass would need to be behind the LMC to produce a similar drag force as the wake. The middle row in Figure \ref{fig:wake_masses} lists the DF acceleration during the same time frames as the top panel, taken from Figure \ref{fig:drag}. 

The distances at which an object of the wake mass would produce a gravitational acceleration equivalent to DF are shown in the bottom row of panels in Figure \ref{fig:wake_masses}. In the Fiducial case, the distances all agree, and are approximately 100 kpc. The Infall distances are roughly 135 kpc, and also show agreement between each DM model. 

In summary, we see that the Fiducial wake acts like an additional LMC-mass object that trails the LMC at a distance of 100 kpc, while the Infall wake is equivalent to an object with roughly $10\%$ the mass of the LMC trailing at a distance of 135 kpc. Additionally, this behavior is insensitive to the assumption of CDM or FDM. 

\begin{table*}
    \centering
    \caption{A non-comprehensive list of recent FDM particle mass constraints in the literature. Mass constraints are listed as the region of \textit{allowed} parameter space, in eV. We include a brief note as to the technique used to derive the constraint as well as the reference. Confidence levels for the constraints vary, but are typically at least 2-$\sigma$.} \label{tab:FDM_mass}
    \begin{tabular}{c c c}
        \hline
        \hline
        Allowed values of $m_a$ [eV] & Technique & Reference \\
        \hline
        $< 1.1\times10^{-22}$ & Cusp-core problem solution & \citealt{marsh_axion_2015} \\
        $> 2.9 \times10^{-21}$ & Lyman-$\alpha$ forest & \citealt{armengaud_constraining_2017} \\
        $< 4\times 10^{-23}$ & Internal kinematics of dSph MW satellites & \citealt{gonzalez-morales_unbiased_2017} \\
        $> 3.75\times10^{-21}$ & Lyman-$\alpha$ forest & \citealt{irsic_first_2017} \\
        $> 10^{-21}$ & Lyman-$\alpha$ forest & \citealt{kobayashi_lyman-alpha_2017} \\
        $> 1.5\times10^{-22}$ & Stellar stream heating & \citealt{amorisco_first_2018} \\
        $< 2\times10^{-20} OR > 8\times10^{-19}$ & Soliton gravity measurements using M87* & \citealt{bar_looking_2019} \\
        $> 4\times 10^{-22}$ & Soliton gravity measurements using Sag A* & \citealt{bar_looking_2019} \\
        $> 6 \times 10^{-23}$ & MW disk star heating & \citealt{church_heating_2019} \\
        $>10^{-18}$ & Core-halo \& BH-halo mass relations & \citealt{desjacques_axion_2019} \\
        $> 5.2 \times 10^{-21}$ & Subhalo mass function via lensing $\&$ stellar streams & \citealt{benito_implications_2020} \\
        $>6 \times 10^{-22}$ & UFD density profiles & \citealt{safarzadeh_ultra-light_2020} \\
        $> 2.1 \times 10^{-21}$ & Subhalo mass function via lensing $\&$ stellar streams & \citealt{schutz_subhalo_2020} \\
        $> 2.2 \times 10^{-21}$ & Stellar stream heating $\&$ MW satellite counts & \citealt{banik_novel_2021} \\
        $> 2\times10^{-20}$ & Lyman-$\alpha$ forest & \citealt{rogers_strong_2021} \\
        $> 3\times10^{-19}$ & Internal kinematics of UFDs & \citealt{dalal_not_2022} \\
        $> 10^{-23}$ & \textit{Planck} $\&$ Dark Energy Survey Year-1 shear measurements & \citealt{dentler_fuzzy_2022} \\
        \hline
        \hline
    \end{tabular}
\end{table*}

\section{Discussion: Simulation Parameters} \label{sec:caveats}

In this section, we explore how our results are affected by changing certain assumptions in our simulation setup. We assess the importance of the FDM particle mass to our results in $\S$ \ref{subsec:FDM_mass}. $\S$ \ref{subsec:LMC_mass} discusses the impact of the uncertainty in the LMC's mass on our observational predictions. We quantify the effect of the stellar halo's velocity dispersion in $\S$ \ref{subsec:sig_star} and discuss implications for the wake's impact on cold stellar substructures. Finally, we discuss the prospects for using the wake to constrain alternative DM models beyond FDM in $\S$ \ref{subsec:SIDM}. To study each of these effects, we run additional simulations which are summarized in Tables \ref{tab:adtl_sims} and \ref{tab:lmc_models}. 

\subsection{The Effect of FDM Particle Mass} \label{subsec:FDM_mass}

As the behavior of FDM is strongly dependent on the particle mass $m_a$, it is important to place our choice of $m_a = 10^{-23}$ eV into context within the literature and test the extent to which a different choice would affect our results. Table \ref{tab:FDM_mass} compiles a list of recent papers which report a constraint on $m_a$ through an astrophysical technique (see also \citealt{ferreira_ultra-light_2021} for a recent review, and Figure 1 of \citet{dome_cosmic_2023} for a graphical approach). We do not guarantee that this list is exhaustive, nor do we include constraints from laboratory or direct-detection experiments. Nevertheless, we hope to demonstrate that FDM particle mass constraints are abundant and may be derived with a very wide range of methods. Notably, the constraints we list here span the entire range of FDM masses ($10^{-26}$ - $10^{-16}$ eV), though almost all come with caveats. 

One common method of constraining $m_a$ relies on trying to detect soliton density cores in dwarf galaxies (e.g. \citealt{bar_looking_2019}; \citealt{desjacques_axion_2019}; \citealt{safarzadeh_ultra-light_2020}). The widest constraint comes from \citet{safarzadeh_ultra-light_2020}, who report that a single-component FDM is incompatible with the observed differences between Fornax and Segue 1's central density profiles. This result relies heavily on the measurement of the Ultra-Faint Dwarf (UFD) density profile slopes, and relaxing the core profile slope constraint from \citet{walker_method_2011} results in a lower bound of $m_a >6 \times 10^{-22}$ eV. Moreover, \citet{chan_diversity_2022} report that there is significant scatter in the FDM soliton core-halo mass relation, which may weaken constraints derived by examining DM density profiles.

Meanwhile, there is a growing tension between the requirements that FDM is light enough that it produces sufficiently large cores in dwarf galaxies (\citealt{marsh_axion_2015}; \citealt{gonzalez-morales_unbiased_2017}) and heavy enough that it is consistent with the small-scale matter power spectrum as inferred from the Lyman-$\alpha$ forest \citep{armengaud_constraining_2017, irsic_first_2017, kobayashi_lyman-alpha_2017, rogers_strong_2021} and other cosmological probes \citep[e.g.][]{dentler_fuzzy_2022}. Such cosmological probes of the FDM mass typically rely on comparisons to simulations performed with traditional \textit{N}-body codes which modify the linear power spectrum of the initial conditions (and sometimes the transfer function) to match that expected of FDM using \texttt{axionCAMB} \citep{hlozek_future_2017}. \citet{schive_contrasting_2016} argue that this approach is valid for power spectrum modeling, though such methods do not consider non-linear effects. Large-scale (box sizes of $L \gtrsim 10 $ Mpc/$h$) cosmological simulations with full SP solvers are becoming available (\citealt{may_structure_2021}; \citealt{may_halo_2023}), which could be used to test the Lyman-$\alpha$ results with higher-fidelity simulations. 

Other methods (such as this work) rely on examining the gravitational effect of FDM granules and/or subhalos on luminous matter. There is a growing class of papers that examine dynamical heating of stars by FDM substructures as a method of placing upper limits on $m_a$ (e.g. \citealt{amorisco_first_2018}; \citealt{church_heating_2019}; \citealt{benito_implications_2020}; \citealt{schutz_subhalo_2020}; \citealt{banik_novel_2021}). Again, none of these studies utilizes a fully self-consistent numerical treatment of the FDM, instead approximating granules as massive extended particles or utilizing only the subhalo mass function. 

A very recent study by \citet{dalal_not_2022} provides one of the more stringent kinematic constraints of $m_a > 3\times10^{-19}$ by examining the heating of stars by granules in the Segue 1 and 2 UFDs. Their simulation technique, outlined in \citet{dalal_dont_2021}, approximates FDM granules as linear perturbations to a static potential. This results in a computationally inexpensive, relatively accurate treatment of the wave behavior of FDM in the idealized case of a spherically symmetric, equilibrium halo. 

\begin{figure}
    \centering
    \includegraphics[width=1.0\columnwidth]{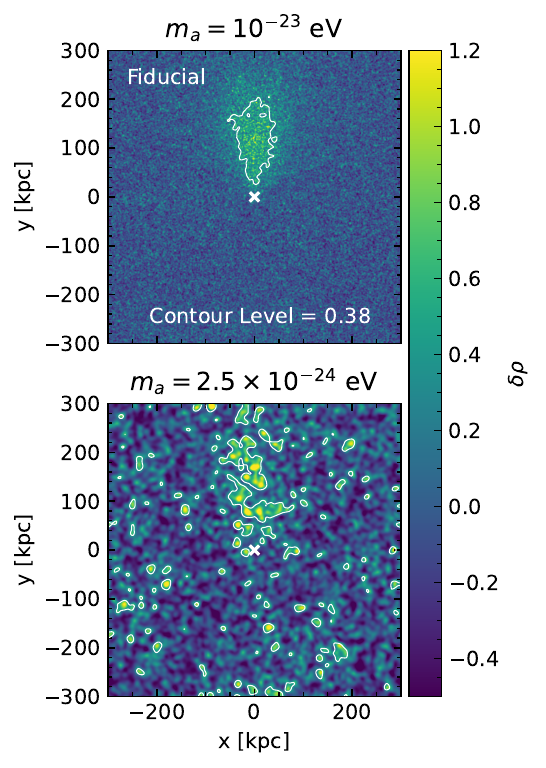}
    \caption{Density of the wakes in both FDM simulations with the Fiducial DM wind parameters (similar to Figure \ref{fig:int_dm_density}). The top panel shows the simulation with $m_a=10^{-23}$ and is identical to the right panel of Figure \ref{fig:int_dm_density}. The bottom panel shows the simulation with $m_a=2.5 \times 10^{-24}$. Decreasing the FDM particle mass increases the de Broglie wavelength, granule size and strength, and acts to wash out the wake slightly. }
    \label{fig:FDM_dens}
\end{figure}

\begin{figure}
    \centering
    \includegraphics[width=1.0\columnwidth]{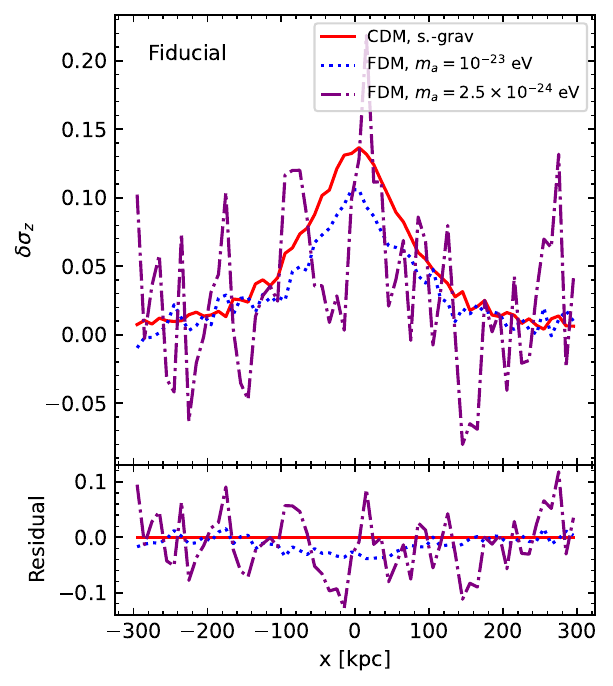}
    \caption{Profile of the $z$-velocity dispersion across the wake (same as Figure \ref{fig:int_dm_radDisp_profile}) with the $m_a = 2.5 \times 10^{-24}$ eV FDM simulation included as the purple dash-dotted line. For clarity, we omit the CDM simulation without self-gravity. The increased granule strength at lower particle mass is also seen in the velocity dispersion. The four-fold decrease in particle mass manifests as a roughly four-fold increase in the amplitude of the oscillations in the low-mass FDM profile. On average, the two FDM profiles show a similar $\delta \sigma_z \sim 0.03$ drop in the peak dispersion compared to CDM. }
    \label{fig:FDM_radDisp}
\end{figure}

\begin{figure}
    \centering
    \includegraphics[width=1.0\columnwidth]{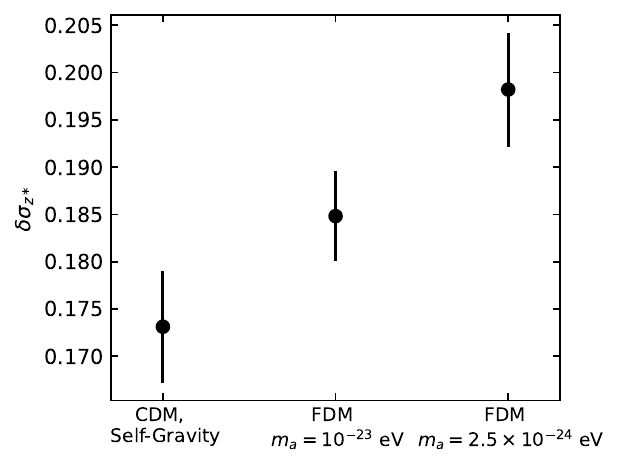}
    \caption{Time-averaged median $z$-velocity dispersion enhancement of the stellar wake (same as Figure \ref{fig:int_stars_radDisp_tavg}) including the CDM simulation with self-gravity and both FDM simulations. The velocity dispersion of the stellar wake is increased more compared to CDM when the FDM particle mass is lower. This demonstrates that the strength of granule heating increases as the FDM particle mass decreases.}
    \label{fig:FDM_radDisp_tavg}
\end{figure}

While very few of these existing constraints have been confirmed with self-consistent, non-linear SP simulations, our choice of $m_a=10^{-23}$ eV is clearly inconsistent with a wide range of observational probes. As justified in Section \ref{subsec:FDM_sim}, this is the largest mass we can feasibly simulate, so it is important to explore how our results are affected by another choice of $m_a$. To determine this, in addition to our $m_a=10^{-23}$ eV simulations, we have performed another set of simulations with $m_a = 2.5 \times 10^{-24}$ eV (see Table \ref{tab:adtl_sims}). We discuss each of our FDM-specific results and their dependence on $m_a$ in turn:

\textbf{Dark Matter Wake Structure:} Figure \ref{fig:FDM_dens} shows the overdensity projections of both FDM simulations with the Fiducial wind parameters (similar to Figure \ref{fig:int_dm_density}). Reducing the mass by a factor of four correspondingly increases the de Broglie wavelength of the FDM particles by a factor of four. As expected, this increases both the size and relative strength of the granule density fluctuations, with peak granule densities within the wake reaching overdensities of $\sim 2.4$ (1.7) for the lower (higher) particle mass. In the low-mass case, some of the background granules (those outside the wake) reach higher overdensities than the half-max of the CDM wake with self-gravity. At higher masses than we are able to simulate, the granules would decrease in size and strength and the density field of the wake would approach the behavior of CDM.

\textbf{Dark Matter Wake Velocity Dispersion:} In Figure \ref{fig:FDM_radDisp}, we reproduce Figure \ref{fig:int_dm_radDisp_profile} with the inclusion of our $m_a = 2.5 \times 10^{-24}$ eV
FDM simulation as a dash-dotted, purple line. The CDM simulation without self-gravity is removed for clarity. The increased de Broglie wavelength of the low-mass simulation causes larger velocity granules, which can be seen as the increased oscillation amplitude in the profile of the low-mass FDM wake. This roughly four-fold increase in the oscillation strength is inversely proportional to the decrease in mass compared to the primary (higher) FDM mass. Despite the oscillations, the two particle masses we consider here show very similar overall/averaged behavior, i.e. when comparing the two masses tested here, our result that the dispersion enhancement of an FDM wake is $\sim 80\%$ that of CDM is unchanged. We caution that this result may not hold for higher particle masses, especially as FDM phenomenology approaches CDM when $m_a$ increases. It is, however, suggestive that the kinematic signatures of FDM wakes are less sensitive to $m_a$ than their density field signature.

\textbf{Kinematics of the Stellar Response:} In $\S$ \ref{sec:star_wakes}, we argued that FDM granule heating is responsible for raising the velocity dispersion of the stellar wake in an FDM universe compared to a CDM universe. 
Following an argument similar to that of \citet{dalal_not_2022}, we can roughly estimate the extent to which granule heating is expected to operate within our windtunnel simulations: FDM granules are approximated as objects of mass $\delta M \approx \bar{\rho} r^3$, where we assume that the granule overdensity fluctuation is of order unity, and the granule radius $r \approx \hbar/m_a \bar{\sigma}$ is set by the de Broglie wavelength associated with the FDM velocity dispersion. Thus, the FDM granules will cause a perturbation in the gravitational potential $\delta \Phi \approx G \delta M / r = G \bar{\rho} r^2$. Stars that encounter granules at a relative velocity of $\sim \bar{\sigma_*}$ will have their velocities perturbed by $\delta v \approx \delta \Phi / \bar{\sigma_*} = G \bar{\rho} r^2 / \bar{\sigma_*}$. Repeated encounters would increase the velocity dispersion of the stars by $\Delta \sigma_* \approx \sqrt{N \delta v^2}$, where $N \approx \bar{\sigma_*} t / r$ is the number of star-granule encounters during a time $t$. Putting all of this together gives
\begin{equation} \label{eqn:granule_heating}
    \Delta \sigma_* \approx \sqrt{\frac{G^2 \bar{\rho}^2 t}{\bar{\sigma_*} \bar{\sigma}^3} \left( \frac{\hbar}{m_a}\right)^3} .
\end{equation}

Notably, Equation \ref{eqn:granule_heating} is derived assuming a uniform density and velocity dispersion of both DM and stars, i.e. similar to our initial conditions. In addition to the increase in granule density within the wake, \citet{lancaster_dynamical_2020} and \citet{vitsos_dynamical_2023} have demonstrated that FDM wakes grow additional interference fringes during the interaction with length scales set by the de Broglie wavelength associated with the wind velocity. Therefore, we do not necessarily expect Equation \ref{eqn:granule_heating} to hold for our simulations but it illustrates that we may expect granule heating to become stronger for lower values of the FDM particle mass.

Figure \ref{fig:FDM_radDisp_tavg} reproduces Figure \ref{fig:int_stars_radDisp_tavg} but includes the low-mass FDM simulation in place of the CDM simulation without self-gravity. The leftmost two points are the same as the rightmost two points in Figure \ref{fig:int_stars_radDisp_tavg}. The stellar wake's velocity dispersion is increased more by the lower-mass FDM wake when compared to CDM, confirming that granule heating becomes stronger when the FDM particle mass decreases. This demonstrates that future observations of the stellar wake's velocity dispersion may be used to place an independent constraint on $m_a$. 

\begin{figure*}
    \centering
    \includegraphics[width=1.0\textwidth]{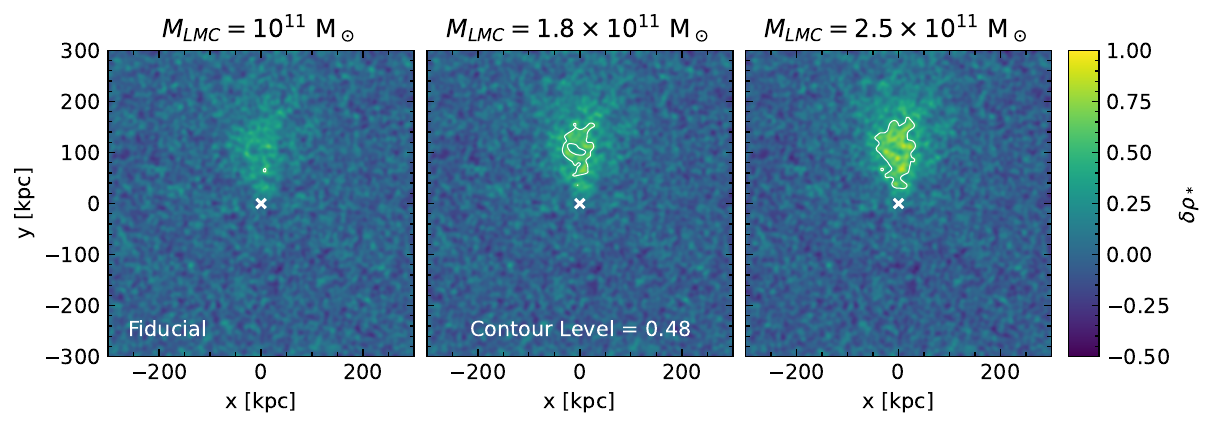}
    \caption{Density of the stellar wake in the three simulations without CDM self-gravity with different LMC mass models (see Table \ref{tab:lmc_models}), computed identically to Figure \ref{fig:int_stars_density}. From left-to-right, the panels show the simulation with the Light LMC, Fiducial LMC (i.e. this panel is identical to the left panel of Figure \ref{fig:int_stars_density}), and the Heavy LMC. Increasing the LMC mass increases the length of the response. However, raising the LMC's mass from $1.8 \rightarrow 2.5\times10^{11}$ M$_\odot$ increases the wake's length by only $\sim$ 25 kpc, compared to the $\sim$50 kpc increase caused by the addition of the DM wake's gravity (see Figure \ref{fig:int_stars_density}). 
    } 
    \label{fig:lmc_mass}
\end{figure*}

Overall, we find that our choices of $m_a$ do not affect our result that an FDM wake is $\sim 20\%$ colder than a comparable CDM wake. We cautiously suggest that these results may hold at higher values of $m_a$, but emphasize the need for higher-resolution simulations conducted with values of $m_a$ that are permitted by other astrophysical constraints to verify this conclusion. Additionally, we should expect granule heating of the stellar wake to decrease as $m_a$ increases, and vice-versa. 

\subsection{The Effect of the LMC's Mass} \label{subsec:LMC_mass}

In $\S$ \ref{sec:star_wakes}, we argued that the length of the stellar wake could be used to reveal the presence of the DM wake, as the DM wake's self-gravity enables the stellar wake to persist for longer than without self-gravity. However, the LMC's mass will affect the strength and length of the stellar wake in a manner that could be degenerate with the presence of the DM wake. To investigate this possibility, we ran two additional simulations with alternative LMC models (see Tables \ref{tab:adtl_sims} and \ref{tab:lmc_models}) of different masses. Both of these additional simulations are run in CDM without DM self-gravity to assess whether a more or less massive LMC could cause a density enhancement in the stellar wake similar to that caused by the addition of the DM wake's gravity. 

Figure \ref{fig:lmc_mass} compares the density of the stellar wakes (similar to Figure \ref{fig:int_stars_density}) in all three simulations without self-gravity. The LMC mass differs between each column, and increases left-to-right. To compare the density response in these simulations to that expected with CDM self-gravity, the contours are set at the half-maximum of the wake's density \textit{with} DM self-gravity, i.e. at the same level as in Figure \ref{fig:int_stars_density}. Increasing the LMC mass increases the strength and length of the response. The wake produced by the Light LMC (left) barely reaches an overdensity of 0.48, while the wake produced by the Heavy LMC is $\sim$ 25 kpc longer than that produced by the Fiducial LMC. 

Therefore, we see that the LMC mass is mildly degenerate with the presence of the DM wake for increasing the length of the stellar wake. However, in $\S$ \ref{sec:star_wakes} we showed that the DM wake's gravity lengthens the stellar wake by $\sim 50$ kpc, roughly twice the increase resulting from raising the LMC mass to $2.5\times10^{11}$ M$_\odot$. Additional constraints on the LMC's mass will also help break this degeneracy. For example, a different assumption for the mass of the LMC will change its orbit \citep[e.g.][]{gomez_and_2015, patel_orbits_2017}. Characterizing the location and kinematics of the wake will constrain the LMC's orbit and in turn its mass \citepalias{garavito-camargo_hunting_2019}. Thus, both the length and location of the wake could be used to break the degeneracy between DM gravity and the LMC's mass for a given MW model. 

\subsection{The Effect of the Stellar Velocity Dispersion} \label{subsec:sig_star}

\begin{figure}
    \centering
    \includegraphics[width=1.0\columnwidth]{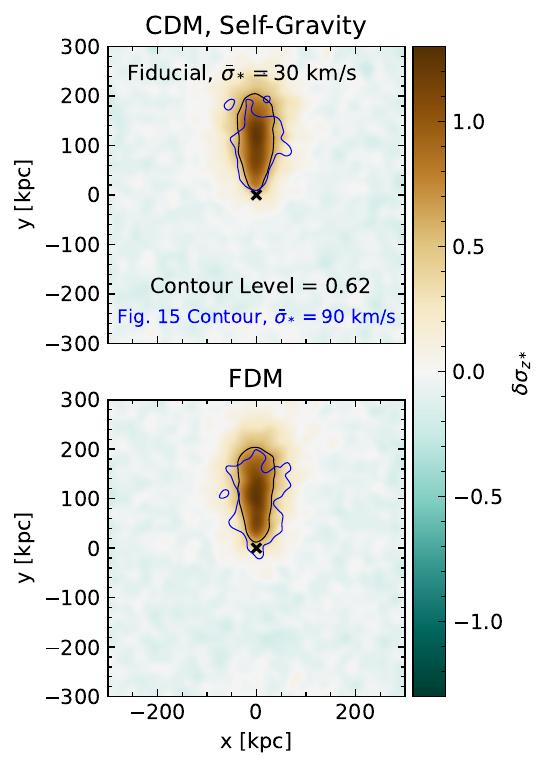}
    \caption{$z$-velocity dispersion of the stellar wake for the simulations with a low stellar velocity dispersion ($\bar{\sigma}_* = 30$ km/s), similar to Figure \ref{fig:int_stars_radDisp}. The response is similar in both DM models. The black contours enclose the region in which the velocity response is greater than half its maximum in CDM, while the blue contours show the wake boundaries from Figure \ref{fig:int_stars_radDisp} for comparison (in which the initial stellar velocity dispersion is 90 km/s). The wake is $\sim 60$ kpc narrower in $x$ when the stellar velocity dispersion is reduced, though the length of the wake does not change significantly.
    }
    \label{fig:int_stars_radDisp_lowsig}
\end{figure}

\begin{figure}
    \centering
    \includegraphics[width=1.0\columnwidth]{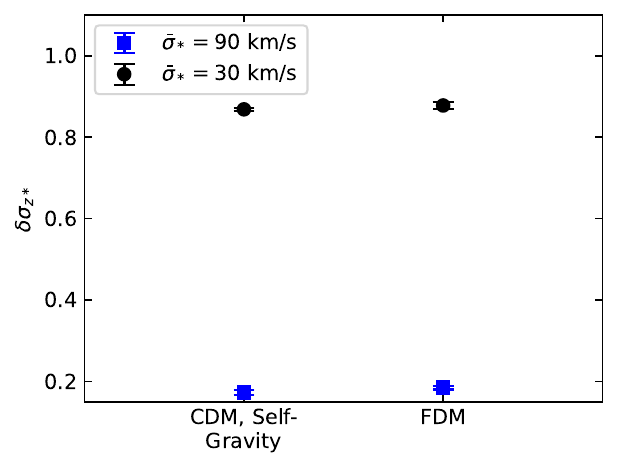}
    \caption{Time-averaged median $z$-velocity dispersion enhancement of the stellar wake as a function of DM model and stellar wind dispersion (similar to Figure \ref{fig:int_stars_radDisp_tavg}). The blue squares show the Fiducial simulations, while the black circles show the simulations with a lowered stellar velocity dispersion. The strength of the response increases by roughly a factor of 5 when the initial stellar velocity dispersion is lowered by a factor of 3. Meanwhile, the strength of granule heating in the FDM simulations is unaffected by the initial velocity dispersion of the stars, i.e. $\delta\sigma_{z*}$ is $\sim 0.01$ higher in FDM compared to CDM in all cases. 
    }
    \label{fig:avg_radDisp_lowstarsig_tAvg}
\end{figure}

While we have so far assumed that the MW stellar halo is smooth and isotropic beyond 70 kpc from the Galactic center, the stellar halo at these distances is likely highly substructured, consisting of streams, shells, and other partially phase-mixed debris from the MW's past accretion events \citep[e.g. see the review by][and references therein]{helmi_streams_2020}. These substructures have lower local velocity dispersions than the phase-mixed component of the stellar halo and will complicate measurements of the wake's influence on halo stars \citep{cunningham_quantifying_2020}.

While we leave a detailed study of the LMC and wake's impact on cold substructure for future work, we can ask whether the differences between CDM and FDM due to granule heating become more pronounced if the stars have an initially low velocity dispersion. To this end, we re-run our Fiducial simulations in CDM and FDM with DM self-gravity and a stellar velocity dispersion of $\bar{\sigma}_* = 30$ km/s (see Table \ref{tab:adtl_sims}), a factor of three lower than in our Fiducial simulations.

Figure \ref{fig:int_stars_radDisp_lowsig} shows the $z$-velocity dispersions of the stellar wakes in these new simulations. The black contour is set at the half-max of the CDM simulation with low stellar dispersion. To compare the size of the wake with the Fiducial simulations (in which $\bar{\sigma}_* = 90$ km/s), we reproduce the wake boundary contours from \ref{fig:int_stars_radDisp} in blue. The impact of lowering the initial stellar velocity dispersion is to narrow the wake (by $\sim 60$ kpc) in both CDM and FDM. 

In Figure \ref{fig:avg_radDisp_lowstarsig_tAvg} we compare the time-averaged median $z$-velocity dispersion within the stellar wakes in our Fiducial simulations to those with a lowered initial stellar dispersion. The strength of the response increases roughly five-fold in the low-stellar-dispersion simulations compared to the Fiducial simulations. While granule heating is still present in the low-stellar-dispersion simulations, i.e. the stellar wake is hotter in FDM than CDM, the difference between CDM and FDM is $\delta\sigma_{z*}\sim 0.01$, identical to the Fiducial simulations.

These results suggest that the LMC and its wake will leave much stronger kinematic signatures in cold stellar substructures compared to phase-mixed populations of stars. These results warrant further testing with simulations of the LMC's infall through a substructured stellar halo.

\subsection{Self-Interacting Dark Matter}\label{subsec:SIDM}
Given the differences we find in the LMC's DM and stellar wakes between FDM and CDM, it is also worth asking whether other DM particle candidates might also impact the DM and/or stellar wake in unique ways. In particular, self-interacting DM (SIDM) (\citealt{carlson_self-interacting_1992, spergel_observational_2000}; see also \citealt{tulin_dark_2018} and \citealt{adhikari_astrophysical_2022} for recent reviews), in which the DM particle has some non-negligible cross section for self-scattering, has emerged as another promising alternative to CDM. In the context of DF wakes, self-scattering between DM particles could potentially alter both the density and velocity of DM particles within the wake, as well as induce a bow shock or mach cone in the DM \citep{furlanetto_constraining_2002}. 

For a constant cross section of $\sigma_\chi = 1$ cm$^2$/g, the mean free path of an SIDM particle within the wake (i.e. at $\sim$ twice the Fiducial wind density of $1.083 \times 10^5$ M$_\odot$/kpc$^3$) is $\sim$45 Mpc, so we do not expect the wake itself to differ from CDM in this case. However, in SIDM the LMC's DM halo would subject to ram pressure from the MW's halo. At the central density of our LMC model's halo ($\sim 7 \times 10^9$ M$_\odot$/kpc$^3$), the mean free path for an SIDM particle with $\sigma_\chi = 1$ cm$^2$/g is reduced to 0.67 kpc, where the scattering may have a non-negligible effect. 

Complicating matters further, velocity-dependent cross sections are compatible with a much wider range of astrophysical observations than constant cross sections \citep[e.g.][]{yoshida_weakly_2000, vogelsberger_subhaloes_2012, vogelsberger_ethos_2016, correa_constraining_2021}. Such velocity-dependent cross sections are typically smaller for larger relative velocities, which would alter the efficacy of SIDM ram pressure as the LMC's orbital velocity changes during its infall. A detailed study of the effects of different SIDM cross-sections on the LMC's wake and DM halo along its orbit would require representing the LMC with a live halo of \textit{N}-body particles, which we leave to future work. 

\section{Conclusions} \label{sec:conclusion}

In this paper, we have presented a suite of windtunnel-style simulations of the LMC's DF wake. Our simulation suite compares the wake at two different points in the LMC's orbit (223 and 70 kpc from the MW), and with three different assumptions for the DM model (CDM with and without self-gravity, and FDM). We also explored the impact of the LMC alone and the LMC plus the DM wake on the MW's stellar halo using the three different DM models. 

Our goals were to quantify the impacts of self-gravity and the DM particle assumption on the DM wake's structure and kinematics. We also sought to determine the response of the stellar halo both with and without the gravity of a DM wake, whether different DM particles leave different signatures in the stellar wake, and if these differences are observable when considering typical observational errors. 

We summarize our conclusions about the DM wakes below:

\begin{enumerate}

    \item \textbf{The FDM and CDM (with self-gravity) wakes both reach comparable peak densities of $\sim 1.6$ times higher than the background.} 

    \item \textbf{The inclusion of self-gravity increases the density of the CDM wake, and extends its length. The self-gravity of the DM wake cannot be ignored.} 
    The inclusion of self-gravity increases the peak overdensity of the wake by $\sim 10$\%, in agreement with \citet{rozier_constraining_2022}. In addition, the LMC DM wake sustains a density that is a factor of 1.38 times larger than the background over a distance $\sim 50$ kpc larger than if self-gravity is ignored. The impact of self-gravity on the properties of the wake is dependent on the LMC's orbital properties and is maximized after the LMC falls within 100 kpc of the Galactic center. At larger distances, the LMC is moving at lower speeds. As such,  particles spend more time under the influence of the LMC's gravity, which reduces the relative contribution of the wake's gravity to its structure. This suggests that the best possible region to search for the influence of the DM wake's gravity observationally is at Galactocentric distances of 70-100 kpc (which also avoids contamination from the Clouds themselves). 
    
    \item \textbf{In FDM, the DM wake is more granular as the DM background grows stochastic interference patterns that interact with the LMC.} While individual granules can reach much higher overdensities than are seen in the CDM wake, the overall density of the FDM wake is similar to CDM when self-gravity is included. 

    \item \textbf{The dispersion of the CDM wake with self-gravity is $\sim1.13$ times higher than the mean dispersion}. The inclusion of self-gravity increases the velocity dispersion of the CDM wake by $\sim 20$\%.
    
    \item \textbf{FDM wakes are $20\%$ colder than CDM wakes regardless of the wind speed and density.} This is due to the reduced response of FDM granules to a steep gravitational potential. Consequently, FDM wakes have a granular structure in kinematic signatures as well (see e.g. Figure \ref{fig:int_dm_radDisp}), compared to the smooth signatures of CDM. This result is insensitive to the FDM particle mass within the range tested here ($m_a = 2.5 \times 10^{-24} - 10^{-23}$ eV), suggesting this result may hold for higher FDM masses.

    \item \textbf{The DF drag forces felt by the LMC are similar in FDM and CDM when self-gravity is included.} This result holds across all simulation parameters that were varied (e.g. wind speed, density: Infall vs. Fiducial case). As such, we do not expect the LMC's orbit to change in an FDM universe compared to a CDM universe. When self-gravity is turned off, the drag force is reduced by $\sim 10\%$, consistent with the $\sim 10\%$ reduction in wake density when self-gravity is removed.
    
    \item \textbf{The LMC's DM wake reaches a mass comparable to the LMC's infall mass in the Fiducial wind case, regardless of the DM model.} To a first approximation, the wake acts like an additional subhalo with a mass of $\sim 1.9 (1.5) \times 10^{11}$ M$_\odot$ when self-gravity is on (off) (comparable to the LMC's infall mass) that trails the LMC by $\sim$ 100 kpc. This implies that the wake is a non-negligible perturber to the dynamics of MW halo tracers.

\end{enumerate}

We summarize our conclusions about the stellar wakes below:

\begin{enumerate}
    \item \textbf{The stellar counterparts to the FDM and CDM (with self-gravity) wakes both reach comparable peak densities of $\sim 1.6$ times higher than the background.} This is similar to the behavior of the DM wakes alone. 

    \item \textbf{The self-gravity of the DM wake causes the stellar wake to peak at higher densities (by $10$\%) and persist over larger distances behind the LMC than if there were no DM wake.} The LMC's gravity will cause the formation of a stellar wake in the absence of DM. However, the stellar wake persists over a larger distance (by $\sim 50$ kpc) if the DM wake self-gravity is included. 

    \item \textbf{In the CDM simulation with self-gravity, the stellar wake velocity dispersion is $\sim 1.173$ times higher than the mean stellar dispersion.} The self-gravity of the DM wake causes the stellar wake's velocity dispersion enhancement to increase by $\sim 5\%$. 
    
    \item \textbf{In the FDM simulations, scattering of stars by FDM granules increases the stellar velocity dispersion relative to the mean by $\sim 5\%$.} Interestingly, this behavior is opposite to that of the DM wake: while the stellar wake is dynamically hotter in FDM, the DM wake is colder in FDM when compared to CDM. The effect of granule heating in the stellar wake will decrease for higher values of $m_a$. 
    
    \item \textbf{Reducing the initial velocity dispersion of the stellar halo by a factor of three (from 90 km/s to 30 km/s) results in an increase in the stellar wake dispersion relative to the mean by a factor of 5 in both the CDM (with self-gravity) and FDM simulations.} This implies that the LMC's wake will have a stronger imprint on the motions of cold substructures in the stellar halo than on phase-mixed halo stars. Meanwhile, the effect of FDM granule heating remains the same when the initial stellar velocity dispersion is lowered. 
    
    \item  \textbf{The angular extent of the stellar wake on the sky can indicate the existence of a DM wake.}  When viewed in Galactic coordinates between distances of 70-100 kpc, the stellar wake appears as an enhancement in the density and radial velocity dispersion of the stellar halo. The response appears in the Galactic southeast, traces the past orbit of the LMC, and extends up to the Galactic plane in $b$ when the DM wake's gravity is included, in agreement with the results of \citetalias{garavito-camargo_hunting_2019}. Without the gravity of the DM wake, the stellar wake decays below an overdensity of 0.34 by $b \approx -20$, and the velocity dispersion enhancement decays below 5.26 km/s by $b \approx -30$. Thus, the length of the stellar wake is an observational sign of the presence of a DM wake, though this is partially degenerate with the LMC's mass. Independent constraints on the LMC's mass or orbit (such as by determining the wake's location) will help break this degeneracy.

    \item \textbf{The differences in the density and velocity dispersion of the stellar wake found across the three models considered (CDM with or without self-gravity, and FDM) persist when the wake is viewed in Galactic coordinates with simulated observational errors, provided at least $10^4$ (900) stars are observed across the sky (in the wake).} Distinguishing FDM from CDM through measurements of the stellar wake's velocity dispersion will be difficult, as the effect of granule heating is only at the percent level. However, our results demonstrate granule heating does play a role in DF wakes and merits further study. Additionally, we find in general that the increased velocity dispersion and extent of the stellar wake is a telltale feature of a DM wake that would distinguish it from cold stellar streams, confirming the findings of \citetalias{garavito-camargo_hunting_2019}. These results underscore the importance of making kinematic measurements when designing observations of the stellar wake. 
    
\end{enumerate}

In this work, we have demonstrated that there are marked differences in the density structure and kinematics of the LMC's DF wake in a CDM vs. an FDM universe, but these differences may be challenging to distinguish in observations of the stellar halo. Significantly, we have also illustrated that the self-gravity of the DM wake plays a crucial role in strengthening and extending the stellar halo's response to the DM wake - providing a new avenue to test for the existence of DM. Next-generation spectroscopic surveys like DESI \citep{desi_collaboration_desi1_2016, desi_collaboration_desi2_2016}, LSST/Vera Rubin Observatory \citep{ivezic_lsst_2019}, and the \textit{Nancy Grace Roman Space Telescope} are poised to provide precision radial velocity and distance measurements of increasing numbers of stars in the stellar halo. These measurements will provide an unprecedented window into the underlying DM distribution of our Local Group \citep{dey_desi_2023}, and potentially the nature of DM itself.

%% Also note that the akcnowlodgment environment does not support long amounts of text. If you have a lot of people and institutions to acknowledge, do not use this command. Instead, create a new \section{Acknowledgments}.
\section{acknowledgments}
H.R.F. would like to thank Peter Behroozi, Arjun Dey, and Dennis Zaritsky for productive discussions that greatly improved this manuscript, Tomer Yavetz for a particularly helpful discussion on the correspondence between CDM and FDM halos, and the anonymous referee for insightful comments that improved the clarity of the paper and presentation of our results.

This work is based upon High Performance Computing (HPC) resources supported by the University of Arizona TRIF, UITS, and Research, Innovation, and Impact (RII) and maintained by the UArizona Research Technologies department. H.R.F. thanks Derrick Zwickl and Chris Reidy for their assistance with MPI troubleshooting, which was made possible through University of Arizona Research Technologies Collaborative Support program. 

We respectfully acknowledge the University of Arizona is on the land and territories of Indigenous peoples. Today, Arizona is home to 22 federally recognized tribes, with Tucson being home to the O’odham and the Yaqui. Committed to diversity and inclusion, the University strives to build sustainable relationships with sovereign Native Nations and Indigenous communities through education offerings, partnerships, and community service.

H.R.F. and G.B. are supported by NSF CAREER AST-1941096.
This work was performed under the auspices of the U.S. Department of Energy by Lawrence Livermore National Laboratory under contract DE-AC52-07NA27344.
E.C.C. acknowledges support for this work provided by NASA through the NASA Hubble Fellowship Program grant HST-HF2-51502.001-A awarded by the Space Telescope Science Institute, which is operated by the Association of Universities for Research in Astronomy, Inc., for NASA, under contract NAS5-26555.
F.A.G. acknowledges support from ANID FONDECYT Regular 1211370 and by the ANID BASAL project FB210003. F.A.G. acknowledges funding from the Max Planck Society through a “Partner Group” grant.
C.F.P.L. acknowledges funding from the European Research Council (ERC) under the European Union’s Horizon 2020 research and innovation programme (grant agreement No. 852839).


\vspace{5mm}
% \facilities{facilites}


\software{
\texttt{Arepo} \citep{springel_e_2010};
\texttt{Astropy} (\citealt{astropy:2013}; \citealt{astropy:2018}; \citealt{astropy:2022});
\texttt{BECDM} \citep{mocz_galaxy_2017, mocz_galaxy_2020};
\texttt{Gadget4} \citep{springel_simulating_2021};
\texttt{Healpy} \citep{Gorski_2005_HEALPix, Zonca_2019_healpy};
\texttt{H5py} \citep{collette_2014_h5py};
\texttt{Matplotlib} \citep{hunter_2007_matplotlib};
\texttt{Numpy} \citep{harris_2020_array};
\texttt{Scipy} \citep{virtanen_2020_SciPy};
          }



\appendix

\section{Comparison of the windtunnel wake to a wake in a live halo} \label{apdx:live_comp}

\begin{figure*}
    \centering
    \includegraphics[width=0.8\textwidth]{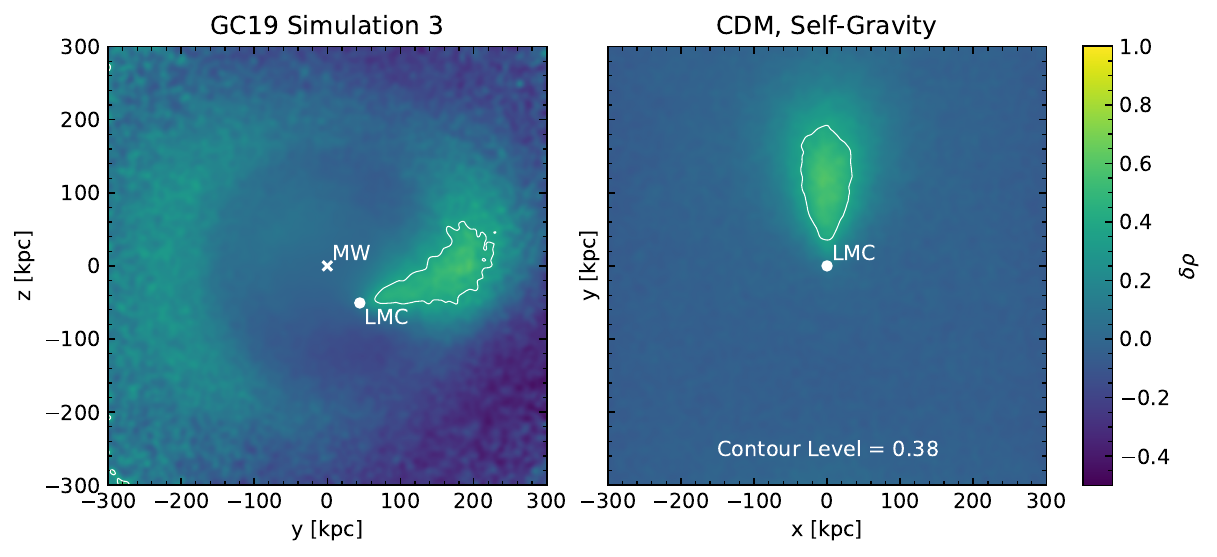}
    \caption{Comparison between the DM overdensity of \citetalias{garavito-camargo_hunting_2019}'s Simulation 3 and our Fiducial CDM simulation with self-gravity. The right panel is a reproduction of the center panel of Figure \ref{fig:int_dm_density}. On the left, we compare this to the overdensity of MW DM particles in \citetalias{garavito-camargo_hunting_2019}'s Simulation 3 when the LMC is at a Galactocentric distance of 70 kpc. We compute the density in the \citetalias{garavito-camargo_hunting_2019} simulation within a 120-kpc-thick slab centered on the Galactic $y$-$z$ plane such that the volume of the slab is identical to our Figure \ref{fig:int_dm_density}. We also draw a contour around the region with an overdensity greater than 0.38 as in the right panel. The size and strength of the wake agree between both panels, showing that the wakes formed in our windtunnels are a good approximation of the wake formed in a live-halo simulation.}
    \label{fig:live_halo_comp}
\end{figure*}

In this Appendix, we compare our Fiducial CDM (with self-gravity) wake to our reference simulation (simulation 3 from \citetalias{garavito-camargo_hunting_2019}). Our windtunnel simulations have several important inherent differences from the full-interaction scenarios presented in \citetalias{garavito-camargo_hunting_2019}: in our simulations, the LMC is represented by a fixed potential instead of as a live halo, there are no Galactic tides from the MW, the LMC ``travels'' in a straight line instead of a curved orbit, and the DM wind parameters are time-independent instead of varying along the orbit as the LMC plunges into the MW's halo. Therefore, it is vital to compare the CDM wake simulated in this paper to the wake in our reference \citetalias{garavito-camargo_hunting_2019} Simulation 3 to ensure that our windtunnel is a reasonable laboratory to study differences between CDM and FDM and the impact of self-gravity on the wake. 

Figure \ref{fig:live_halo_comp} compares the strength and size of the DM wake in \citetalias{garavito-camargo_hunting_2019}'s Simulation 3 (left) to our Fiducial CDM simulation with self-gravity (right). Similar to Figure \ref{fig:int_dm_density}, we draw a contour enclosing the region with an overdensity greater than half the maximum overdensity in the right panel. Both wakes are very similar in strength and size, demonstrating that our windtunnel simulation framework can well-reproduce the wake formed in a full-interaction simulation.

\section{Transformation From Windtunnel to Galactic Coordinates}\label{apdx:coord_trans}

To perform transformations between coordinate frames (windtunnel simulation box, Galactocentric, and Galactic), we make use of \texttt{Astropy} version 4.2.1 (\citealt{astropy:2013}; \citealt{astropy:2018}; \citealt{astropy:2022}), and adopt the definitions and conventions of Galactocentric and Galactic coordinates as given in this version of \texttt{Astropy}. Our steps for transforming between simulation box coordinates and Galactocentric coordinates are as follows:

\begin{enumerate}
    \item For reference, we use the LMC's present-day location and velocity vector from \citet{kallivayalil_third-epoch_2013}. The LMC's orbit, as usual, is taken from \citetalias{garavito-camargo_hunting_2019}'s Simulation 3.
    \item We rotate the simulation box such that the LMC's unit velocity vector (-$\hat{y}$ in the windtunnel frame) points in the same direction as the LMC's velocity vector at $r_{MW} = 70$ kpc from the reference simulation. 
    \item Next, we translate the simulation box such that the center of the box (where the LMC potential is located) matches the present-day location of the LMC. This ensures that the LMC is as close as possible to the correct location on the sky after the next step.
    \item A further translation matches the location of the straight windtunnel orbit and the curved orbit from the reference simulation at a Galactocentric distance of 70 kpc. Together with the rotation, this ensures the LMC's path in the windtunnel is tangent to the LMC's orbit at 70 kpc, which is the location our Fiducial wind parameters are drawn from.
    \item Finally, the particles are given a velocity boost to remove the bulk wind velocity, i.e. to ensure the wind particles have no net motion in a Galactocentric frame. 
\end{enumerate}

Figures \ref{fig:allSky_density}, \ref{fig:avg_obsFrame_density_obsErrs}, \ref{fig:allSky_radVel}, and \ref{fig:avg_obsFrame_radDisp_obsErrs} in $\S$ \ref{subsec:obs} are in Galactic coordinates, and we use \texttt{Astropy}'s built-in functionality for transforming between Galactocentric and Galactic coordinates. When plotting the velocities of stars in Galactic coordinates, we also remove the Sun's motion about the Galactic Center. 


\bibliography{references}{}
\bibliographystyle{aasjournal}



\end{document}